\pdfoutput=1
\documentclass[12pt]{article}
\usepackage[ngerman,english,british]{babel}
\usepackage{epsfig}
\usepackage{lscape}
\usepackage{amsmath,amsfonts,amssymb}
\usepackage{eurosym}
\usepackage{natbib}
\usepackage{url}
\usepackage{setspace}
\usepackage{caption}
\usepackage{booktabs}
\usepackage{subcaption}
\usepackage{arydshln}
\usepackage{rotating}
\usepackage{pdflscape}
\usepackage{multirow}
\usepackage{bm}
\usepackage{graphicx}
\usepackage[hang,flushmargin]{footmisc}

\usepackage{xcolor}

\usepackage[margin=1in]{geometry}

\makeatletter
\def\@sect#1#2#3#4#5#6[#7]#8{\ifnum #2>\c@secnumdepth
     \let\@svsec\@empty\else
     \refstepcounter{#1}\edef\@svsec{\csname the#1\endcsname. \hskip 0.4em}\fi
     \@tempskipa #5\relax
      \ifdim \@tempskipa>\z@
        \begingroup #6\relax
          \@hangfrom{\hskip #3\relax\@svsec}{\interlinepenalty \@M #8\par}%
        \endgroup
       \csname #1mark\endcsname{#7}\addcontentsline
         {toc}{#1}{\ifnum #2>\c@secnumdepth \else
                      \protect\numberline{\csname the#1\endcsname}\fi
                    #7}\else
        \def\@svsechd{#6\hskip #3\relax  
                   \@svsec #8\csname #1mark\endcsname
                      {#7}\addcontentsline
                           {toc}{#1}{\ifnum #2>\c@secnumdepth \else
                             \protect\numberline{\csname the#1\endcsname}\fi
                       #7}}\fi
     \@xsect{#5}}
\makeatother

\makeatletter
\renewcommand{\section}{\@startsection{section}{1}{0mm}{-\baselineskip}{0.25\baselineskip}{
\bfseries
\normalsize
}
}
\renewcommand{\subsection}{\@startsection{subsection}{2}{0mm}{-\baselineskip}{0.25\baselineskip}{\raggedright\normalfont\normalsize\itshape}}
\renewcommand{\subsubsection}{\@startsection{subsubsection}{3}{0mm}{-\baselineskip}{0.25\baselineskip}{\raggedright\normalfont\small\scshape}}
\def\@begintheorem#1#2{\trivlist \item[\hskip \labelsep{\bf #1\ #2:}]\it}
\makeatother

\makeatletter
\def\monthname{\ifcase\month\or
January\or February\or March\or April\or May\or June\or
July\or August\or September\or October\or November\or December\fi}
\makeatother

\renewenvironment{abstract}
 {\begin{center}\normalsize\textsc{Abstract}%
 \end{center}\begin{quote}\normalsize}
 {\end{quote}}

\renewcommand{\appendix}{\footnotesize\parindent 0cm\setlength{\parskip}{\medskipamount}\setcounter{equation}{0}%
\renewcommand{\theequation}{A.\arabic{equation}}}

\newtheorem{proposition}{\small\sc Proposition}
\newtheorem{assumption}{\small\sc Assumption}

\newtheorem{definition}{\small\sc Definition}

\newcommand{\indep}{\perp\!\!\!\!\perp}

\begin{document}

\begin{titlepage}
\vspace*{0.1cm}

\setcounter{page}{0}

  \begin{center}%
    {\Large

Estimating Causal Moderation Effects with Randomized Treatments and Non-Randomized Moderators
\\ \vspace{1.25cm}
\large
\emph{Journal of the Royal Statistical Society: Series A} \\ \vskip 0.15em
(Forthcoming)
\\
 \sc \par}%
       \vskip 3em%
    {\large
     \lineskip .75em%
      \begin{tabular}[t]{c}%
      Kirk Bansak \\
     \end{tabular}\par}%
    \vskip 1.5em%
    {\small 
            \monthname \ \number\year \par}%
      \vskip 5.0em%

  \end{center}\par

\begin{abstract}

\noindent 
Researchers are often interested in analyzing conditional treatment effects. One variant of this is ``causal moderation," which implies that intervention upon a third (moderator) variable would alter the treatment effect. This study considers the conditions under which causal moderation can be identified and presents a generalized framework for estimating causal moderation effects given randomized treatments and non-randomized moderators. As part of the estimation process, it allows researchers to implement their preferred method of covariate adjustment, including parametric and non-parametric methods, or alternative identification strategies of their choosing. In addition, it provides a set-up whereby sensitivity analysis designed for the average-treatment-effect context can be extended to the moderation context. To illustrate the methods, the study presents two applications: one dealing with the effect of using the term ``welfare" to describe public assistance in the United States, and one dealing with the effect of asylum seekers' religion on European attitudes toward asylum seekers.
\end{abstract}

\bigskip
\bigskip

\vspace*{0.35cm}
\vfill
 \footnoterule
    {\footnotesize
    
\noindent Kirk Bansak, Department of Political Science, University of California San Diego, La Jolla, CA 92093. Email: \texttt{kbansak@ucsd.edu}. \\
\noindent 
I thank Justin Esarey, Justin Grimmer, Jens Hainmueller, Guido Imbens, Ken Scheve, Dustin Tingley, Mike Tomz, and Stefan Wager for helpful comments. \vspace*{0.2cm}
\noindent\footnotesize }
\end{titlepage}

\newpage
\setcounter{page}{1} \addtolength{\baselineskip}{0.1\baselineskip}

\section{Introduction}

Experimental researchers are often interested in analyzing conditional treatment effects. By better understanding how the magnitude or sign of a treatment effect may depend upon other variables, researchers can better explain the phenomena they are studying and determine the likelihood that a treatment will be effective if introduced to other populations in the future. Most commonly, researchers will investigate whether the effect of a treatment $T$ on an outcome $Y$ is conditional upon the value of a third variable $S$. 

However, if researchers find evidence of treatment effect conditionality, they then face an additional hurdle in interpreting the nature of that conditionality. On the one hand, they may simply observe a difference in the treatment effect across different values of $S$ because the treatment is more or less impactful for different segments of the population. This phenomenon is often referred to as treatment effect heterogeneity. In this case, it can be established that the treatment effect is different across sub-groups with different values of $S$, but what cannot be established is whether that difference is attributable to $S$ itself.

On the other hand, however, researchers may be interested in further claiming that $S$ is precisely the reason for the difference in treatment effects---that is, changes in $S$ actually cause changes in the effect of $T$ on $Y$. This phenomenon, which is the focus of the present study, will be referred to as the ``causal moderation" of treatment effects, which is the causal effect of a third variable on the treatment's effect on the outcome. In contrast to treatment effect heterogeneity, which deals with extant variation in the treatment effect across the population, causal moderation implies that counterfactual intervention upon a third (moderator) variable would alter the treatment effect. This study examines whether, when, and how experimental researchers can actually investigate causal moderation rather than limit their focus on treatment effect heterogeneity.

Evidence of causal moderation has important theoretical and policy implications that do not follow from the existence of treatment effect heterogeneity in general, and as such, researchers are often interested in making claims about causal moderation. For instance, the existence and magnitude of causal moderation determines the extent to which a treatment effect would be stable, amplified, or attenuated if a third variable were also manipulated, thereby providing unique insights on the potential generalizability of the treatment effect. 

Researchers focused on causal moderation, however, must confront additional estimation and identification hurdles when the moderator variable has not been randomized. This is a common situation in experimental social science, where financial, logistical, physical, or other constraints allow the researcher to randomize certain variables (the treatments) but not randomize other variables that are believed to interact with the treatment. In this situation, treatment effect conditionality is most commonly analyzed on the basis of conventional subgroup analysis and/or underspecified interaction regression models. As this study will show, these standard approaches generally do not provide unbiased or consistent estimates of causal moderation effects given non-randomization of the moderator. Yet applied researchers using these standard approaches will often make claims suggestive of causal moderation without fully grappling with the identification hurdles. For instance, a study on partisan identity and political emotion evaluates the extent to which political messages threatening a party's electoral loss cause anger for copartisans \citep{huddy2015expressive}. In assessing whether the ``partisan threat" experimental treatment leads to greater anger among respondents with stronger partisan identity, the authors describe this as the ``effect of partisan identity...on emotional response to partisan threat and reassurance," suggesting a causal moderation phenomenon. 

This study considers the conditions under which causal moderation can be identified and presents a generalized framework for estimating causal moderation effects in the specific design context of randomized treatments and non-randomized moderators. As part of the estimation process, it allows researchers to implement their preferred method of covariate adjustment, including parametric and non-parametric methods, or alternative identification strategies of their choosing. To illustrate the methods, two applications are also presented: one dealing with the effect of using the term ``welfare" to describe public assistance in the United States and whether that effect is moderated by party identification, and one dealing with the effect of asylum seekers' religion on European attitudes toward asylum seekers and whether it is moderated by nationalism.

\section{Defining Causal Moderation}

\subsection{Set-up and Notation}

Suppose a simple random sample of size $N$ is drawn from a larger population of interest. Units in the sample are indexed by $i = 1,...,N$, and are assigned to either a treatment condition or a control condition. In addition to measuring an outcome for each unit, also assume measurement of a third variable which will be called a moderator. Employing the potential outcomes framework of \cite{neyman1923applications} and \cite{rubin1974estimating} to define the causal effects of interest, this study posits for each unit $i$ the existence of $Y_i(t,s) \in \mathcal{Y}$ at all values of the treatment $t \in \mathcal{T}$ and moderator $s \in \mathcal{S}$. $Y_i(t,s)$ denotes the potential outcome of unit $i$ if that unit had treatment status $t$ and moderator value $s$.

For each unit in the sample, we observe $(Y_i, T_i, S_i, \bm{X}_i)$, where $Y_i \in \mathcal{Y}$ denotes the realized outcome, $T_i \in \mathcal{T}$ denotes the assigned treatment status, $S_i \in \mathcal{S}$ denotes the observed moderator value, and $\bm{X}_i \in \bm{\mathcal{X}}$ denotes a vector of other observed covariates. Note that $S_i$ and $\bm{X}_i$ will be assumed to be pre-treatment. Under the definition of the potential outcomes, the realized outcome $Y_i = Y_i(T_i, S_i)$. Thus, only one potential outcome is observed for each unit. However, implicit in the postulation of potential outcomes $Y_i(t,s)$ for all $t \in \mathcal{T}$ and $s \in \mathcal{S}$ is that both the treatment and moderator are assumed to be mutable at the individual level. The variables contained in $\bm{X}_i$ may or may not be manipulable. 

Note that this study proceeds from the perspective of sampling from a larger population. Hence, the causal estimands that will be presented pertain to the target population, with expectations taken over the population distribution. However, similar to standard frameworks for identifying an average treatment effect (ATE), the general identification results are not exclusive to this particular sampling context and would proceed similarly in both the super-population approach and the sample-as-population approach, where the sample observed is treated as the population of interest \citep[see][]{imai2008misunderstandings, imbens2015causal, aronow2016does}. Indeed, the sample-as-population perspective can be viewed as a special case that nests within the larger-population perspective, which follows from considering a sample of size $N$ (without replacement) from a population of the same size. 

For simplicity, the rest of the study will consider binary versions of the treatment and moderator: $\mathcal{T} \equiv \{0,1\}$ and $\mathcal{S} \equiv \{0,1\}$. As will be presented, the causal estimands of interest will employ comparisons of potential outcomes $Y_i(t,s)$ where $t$ and $s$ take on two separate values at a time, and the values $0$ and $1$ will be used for convenience. In practice, however, the treatment and moderator could take on any number of values, and $0$ and $1$ used to denote any pairs of values of interest.

\subsection{Treatment Effect Heterogeneity vs. Causal Moderation}

To illustrate what is at stake in investigating causal moderation, in contrast to other types of treatment effect conditionality, consider the following hypothetical experiment. Imagine a study where, in order to investigate a potential pathway for increasing local public goods provision ($Y$) across towns (the units of analysis) in a developing country, researchers run a field experiment that draws a sample of towns and then randomly assigns certain towns to a program ($T$) that provides specialized training to its local public officials. The researchers find that on average the treatment works---the training program improves public goods provision---but also that it appears to have differential effects for different types of towns. Specifically, the researchers suspect that the treatment is more effective in towns that have more officials per capita, i.e. towns with higher governmental capacity ($S$). Deciding how to extend their analysis and interpret their results then depends upon the researchers' goals.

As a first goal, the researchers may simply want to find treatment effect heterogeneity. That is, they may want to know the types of places where the training program is most effective, so that a future training policy can be rolled out in a targeted and cost-efficient manner. They do not necessarily need to know why it is more effective in certain places; they just need to be able to predict what those places are likely to be. In this particular example, investigation of treatment effect heterogeneity across towns with low and high capacity levels could proceed using simple subgroup analysis. From this perspective, researchers would be interested in the following estimand, which captures treatment effect heterogeneity:
$$E \big[Y_i(1,1) - Y_i(0,1) | S_i = 1] - E \big[ Y_i(1,0) - Y_i(0,0) | S_i = 0 \big]$$
where the expectations are taken with respect to the distribution of potential outcomes over the population from which the sample was drawn. 

Given that the expectation for each potential outcome is taken conditional on the observed moderator value being the value of the moderator in the potential outcome, this estimand effectively treats the moderator as a purely observational variable with no causal component of its own being measured. In other words, it captures the difference between the average treatment effect (ATE) of $T$ for units whose moderator value is observed to be $1$ and the ATE for units whose moderator value is observed to be $0$. Accordingly, this estimand could be re-expressed using reduced form potential outcomes $Y_i(t)$, $t \in \{0,1\}$, as follows: $E \big[Y_i(1) - Y_i(0) | S_i = 1] - E \big[ Y_i(1) - Y_i(0) | S_i = 0 \big]$. (Note also that for treatment effect heterogeneity, the moderator would not need to be mutable at the individual level.) Researchers may also want to identify other variables across which the treatment effect is heterogeneous but may not know which observed variables to focus attention on. Recent years have seen an increase in interesting research into new methods to discover, estimate, and perform statistical inference on treatment effect heterogeneity beyond simple subgroup analysis, much of which integrates machine-learning techniques \citep{ding2016randomization, athey2016recursive, wager2018estimation, beygelzimerlangford2009, dudiketal2011, fosteretal2011, greenkern2012, imairatkovic2013, suetal2009, zeileisetal2008, weisbergpontes2015, tianetal2014, ratkovic2017sparse, kunzel2017meta, powers2017some}.

In some cases, however, discovering the existence and extent of treatment effect heterogeneity may not be the ultimate goal of a researcher's inquiry into causal effect conditionality. Instead, resarchers may want to estimate causal moderation effects. At stake here is the ability to claim that a moderator variable truly causes changes in the treatment effect, or that the treatment effect would remain stable if the moderator itself were to be manipulated. The corresponding causal moderation estimand of interest will be referred to as the average treatment moderation effect (ATME) and denoted by $\delta$.
\vspace{0.0cm}
\begin{definition} The average treatment moderation effect (ATME) is defined as \addtolength{\baselineskip}{-0.3\baselineskip}
$$\delta = E \big[ \{Y_i(1,1) - Y_i(0,1)\} - \{Y_i(1,0) - Y_i(0,0)\} \big]$$
where the expectation is taken over the population distribution.
\end{definition}
Whereas treatment effect heterogeneity deals with variation in the treatment effect across different spaces of the observed distribution in the population, causal moderation implies that counterfactual intervention upon the moderator variable would alter the treatment effect. 

In the example above, the researchers may have found that the training program is more effective in high-capacity towns, but they did not randomly assign capacity and thus this result could be because capacity is related to any number of other things that are truly responsible for increasing the effectiveness of the treatment. If their interest is in causal moderation, then the researchers need to know if it truly is higher capacity that causes the training program to be more effective. The answer to this question has theoretical and practical policy importance. If there is reason to believe that policy or other shocks in the future may lead to local capacity growth (or loss), then the researchers should be interested to know how this will likely impact the success of training programs that are introduced in the future (e.g. whether the treatment effect will generalize to new scenarios). Alternatively, they may be interested to know if it would be cost-efficient to add funding for additional public official employment to future training program roll-outs (e.g. whether the treatment regime can be optimized via additional interventions). To achieve this, finding treatment effect heterogeneity is not sufficient; what is needed is the estimation of causal moderation.

\subsection{Causal Moderation and Interaction in Other Contexts}

The ATME as defined in this study is an estimand that has been discussed in previous scholarship on variable interactions. \cite{vanderweele2009distinction, vanderweele2015} has referred to this phenomenon as ``causal interaction" or more simply ``interaction" (while referring to treatment effect heterogeneity as ``effect modification") and \cite{egami2019causal} use the term ``average interaction effect." However, these previous studies consider the estimand within different design contexts than that of the present study. \cite{vanderweele2009distinction, vanderweele2015} considers identification and estimation of the estimand in the general case, without explicitly considering whether the variables have been randomized. In contrast, the present study considers the specific case faced by experimental researchers of one randomized variable and one non-randomized variable, providing a more targeted set of identification assumptions that researchers can think through intuitively, as well as a flexible and straightforward estimation strategy for this context. Conversely, \cite{egami2019causal} discuss the estimand in the special case of two randomized treatments. Consequently, their context does not confront the same identification and estimation challenges present in the design context of interest in this study.

Finally, it is important to explicitly note that this study formulates the ATME (as well as treatment effect heterogeneity defined earlier) in terms of a difference model. In certain scholarly contexts, such as the study of risk factors in epidemiology, a key distinction is made between exposure variable interactions measured on a difference scale versus a ratio or other scale \citep[for discussion of this issue and its relationship to the concept of `biologic' versus statistical interactions, see][]{rothman1980concepts, andersen2015competing}. This can be an important distinction to make because an interaction may manifest on one scale but not another. In addition, as has been observed with respect to the difference-in-differences method of estimating treatment effects \citep[e.g.][]{meyer1995natural}, nonlinear transformations of the outcome variable may also affect the results. Throughout its entirety, this study adopts a difference perspective on variable interaction and moderation, as is clear from the `difference-in-differences' form of both the ATME and treatment effect heterogeneity estimands defined above.

\section{Identification and the Parallel Estimation Framework}

This section lays out the assumptions under which the ATME can be identified, presuming a design context involving a randomized treatment and non-randomized moderator, and introduces a generalized framework for estimating the ATME. 

\subsection{Identification}

To achieve identification of the ATME, this subsection presents a strategy in which selection on observables is applied to the moderator. (However, note that other identification strategies could also be applied under alternative assumptions, a matter that will be revisited later.) The first identification assumption is the stable unit treatment value assumption (SUTVA).
\begin{assumption}[SUTVA] \label{assump:sutva} 
\hfill \break If $T_i = T'_i$ and $S_i = S'_i$, then $Y_i(\vec{T},\vec{S}) = Y_i(\vec{T}',\vec{S}')$, where $\vec{T}$ and $\vec{S}$ denote the full treatment and moderator vectors across subjects in the sample $i = 1,2,...,N$.
\end{assumption}

In accordance with the design context of interest, the second identification assumption is that the treatment has been randomized. While both the identification results and estimation approaches that will be presented could accommodate any randomization scheme where all units share the same treatment assignment probability---including Bernoulli randomization, complete randomization, and stratified randomization with constant probabilities across strata---complete randomization simplifies variance derivation and estimation.
\begin{assumption}[Complete Randomization of the Treatment] \label{assump:exog}
\hfill \break 
$P(\vec{T} = a) = P(\vec{T} = a')$ for all $a$ and $a'$ such that $\iota^T a = \iota^T a'$ where $\iota$ is the N-dimensional column vector with all elements equal to one.
\end{assumption}

The third assumption is that the moderator $S$ and a set of observed confounder variables $\bm{X}$ are pre-treatment, and hence the treatment can have no individual-level effects on $S$ or $\bm{X}$. Combined with Assumption \ref{assump:exog}, this implies that the moderator, confounder variables, and potential outcomes are jointly independent of the treatment.
\begin{assumption}[Pre-Treatment Moderator and Confounders] \label{assump:pretre}
\hfill \break 
The moderator $S$ and a set of observed confounder variables $\bm{X}$ are pre-treatment such that, given randomization of the treatment,
$$\big( Y_i(1,1),Y_i(0,1),Y_i(1,0),Y_i(0,0),S_i,\bm{X}_i \big) \indep T_i$$
\end{assumption}

The fourth identification assumption is conditional independence between the potential outcomes and the moderator.
\begin{assumption}[Moderator Conditional Independence] \label{assump:soo}
$$\big( Y_i(1,1),Y_i(0,1),Y_i(1,0),Y_i(0,0) \big) \indep S_i \: \Big| \: \bm{X}_i$$
\end{assumption}
Note that given a randomized treatment and pre-treatment status of $S$ and $\bm{X}$, the identification burden implied by Assumption \ref{assump:soo} is no more stringent than the burden involved in the standard causal identification of an ATE via selection on observables. In addition, identification via selection on observables also requires a common support assumption.
\begin{assumption}[Moderator Common Support] \label{assump:cs}
$$0 < P(S_i=1|\bm{X}_i) < 1$$
\end{assumption}

The assumptions delineated above allow for identification of the ATME. See Appendix A in the Supplementary Materials (SM) for a proof.

\vspace{0.0cm}
\begin{proposition} \label{prop:main} \addtolength{\baselineskip}{-0.4\baselineskip}
Given Assumptions \ref{assump:sutva}-\ref{assump:cs},
\begin{align} \label{eq:prop1}
\delta & = & E_{\bm{X}} \Big[ \:\:  E[Y_i|T_i=1,S_i=1,\bm{X}_i] - E[Y_i|T_i=0,S_i=1,\bm{X}_i] \\
& & - E[Y_i|T_i=1,S_i=0,\bm{X}_i] + E[Y_i|T_i=0,S_i=0,\bm{X}_i] \:\: \Big] \nonumber
\end{align}
where the outer expectation is taken with respect to the population distribution of $\bm{X}$.
\end{proposition}

\subsection{Estimation}

The expression in Proposition \ref{prop:main} provides a preliminary basis for estimation of $\delta$. However, for a number of reasons that are described below, it is useful to disaggregate the expression into two separate components. Specifically, Assumption \ref{assump:pretre} implies that $F_{\bm{X} | T}(\bm{x} | t) = F_{\bm{X}}(\bm{x})$ for all $\bm{x}$ and $t$, where $F_{\bm{X}}$ denotes the joint distribution function of $\bm{X}$ and $F_{\bm{X} | T}$ denotes the joint distribution function conditional on $T$. Thus, $E_{\bm{X}} \Big[E[Y_i|T_i=t,S_i=s,\bm{X}_i] \Big] = E_{\bm{X}|T=t} \Big[E[Y_i|T_i=t,S_i=s,\bm{X}_i] \Big| T_i = t \Big]$ for any $t$, $s$. Hence, equation (\ref{eq:prop1}) can be re-expressed as follows:
\begin{align} \label{eq:parallel}
\delta & = & E_{\bm{X}|T=1} \Big[ \:\: E[Y_i|T_i=1,S_i=1,\bm{X}_i] - E[Y_i|T_i=1,S_i=0,\bm{X}_i] \:\: \Big| \:\: T_i=1 \:\: \Big] \\
& & - E_{\bm{X}|T=0} \Big[ \:\: E[Y_i|T_i=0,S_i=1,\bm{X}_i] - E[Y_i|T_i=0,S_i=0,\bm{X}_i] \:\: \Big| \:\: T_i=0 \:\: \Big] \nonumber
\end{align}
\vspace{0.0cm}
\indent While this may seem a simple mathematical re-expression, seeing the estimand explicitly in this form is valuable and informative for several reasons. First, it points to an initially non-obvious opportunity for estimation in a manner that is highly accessible to applied researchers. Specifically, it highlights that researchers can simply subset their data by treatment level and then proceed to apply covariate-adjustment estimation strategies separately to the treated units' data and control units' data. That is, 
\begin{align} \label{eq:subset}
E_{\bm{X}|T=t} \Big[ \:\: E[Y_i|T_i=t,S_i=1,\bm{X}_i] - E[Y_i|T_i=t,S_i=0,\bm{X}_i] \:\: \Big| \:\: T_i=t \:\: \Big] 
\end{align}
can be estimated separately for $T_i=0$ and $T_i=1$ by a covariate-adjustment estimator of the researcher's choice. This is equivalent to holding $T$ constant and then estimating the causal effect of $S$ on $Y$ as one normally would in an observational study. A variety of straightforward and widely used methods for performing this estimation already exist, such as regression, matching, and propensity score weighting. As a result, applied researchers can easily adapt the tools they are already familiar with and prefer using. The simple difference between the two subsetted estimates then constitutes an estimate of the ATME. While estimation of the ATME in this manner is not strictly necessary, it provides a highly accessible and flexible approach, which will be referred to as the ``parallel estimation framework."

Second, the disaggregation and ensuing parallel estimation approach improves not only accessibility but also tractability in the case of certain nonparametric covariate-adjustment methods. In particular, it facilitates the use of matching, one of the most popular methods used in the estimation of causal effects of non-randomized variables. Again, note that estimation of the ATME via the parallel estimation framework is not strictly necessary; one could instead specify a general model for $E[Y_i | T_i, S_i, \bm{X}_i]$ as the basis for estimating the ATME. Of course, the key caveat for this (and also for specifying subsetted models in the parallel estimation approach) is that the model must be correctly specified. To avoid such model dependence, researchers interested in estimating causal effects involving non-randomized variables often turn to matching \citep[see][]{ho2007matching, stuart2010matching}. Employing matching would also be fruitful in this regard for estimating the ATME. Looking at the expression as presented in Proposition \ref{prop:main}, one would be inclined to match quadruples of observations, in contrast to the normal requirement to match pairs. This would be problematic for a couple of reasons. Seeking to minimize the distance between four units requires minimizing some chosen metric that aggregates six pairwise distances, and hence achieving a certain quality of match (e.g. average pairwise distance) with finite data becomes more difficult with quadruples than with pairs. In addition, as a more practical matter, while there already exist reliable implementations of pair matching that are accessible to applied researchers across a range of popular statistical software, implementations of quadruple matching are not similarly pervasive. Thus, by allowing for the matching of pairs, the parallel estimation framework effectively enables the use of matching in the ATME estimation.

Third, the disaggregation also provides simplifying insights on how to characterize the statistical properties of and conduct inference on estimates of the ATME. As already described above, the parallel estimation framework allows researchers to employ their preferred method of covariate adjustment to estimate the quantity in expression (\ref{eq:subset}) separately for $t = 0,1$. Under the assumptions presented earlier along with any functional form assumptions made by the researcher to accompany their estimation method (e.g. linearity in the case of employing linear regression), the estimates of expression (\ref{eq:subset}) for $t = 0,1$ will exhibit the unbiasedness and/or consistency of the estimation method. In turn, given the difference form, the estimate of the ATME will inherit this unbiasedness and/or consistency. Furthermore, the two subsets ($T_i=0$ and $T_i=1$) can be viewed as approximately independent. As a result, the variance of the ATME estimate can be approximated and estimated directly as the sum of the variances of the two within-subset estimates. In addition, the parallel estimation framework also facilitates a sensitivity analysis of the ATME estimate, described in detail in Appendix D.

Finally, the fourth benefit of the disaggregation is that it helps to formally reveal how certain common approaches used by researchers in the analysis of treatment effect conditionality are not well-suited to estimating the ATME. As described, estimation of expression (\ref{eq:subset}) is performed separately for $t=0,1$. In contrast, however, subsetting the units by moderator level and taking the difference between estimates of the treatment effects within those subsets would lead to biased and inconsistent estimates of the ATME, as $F_{\bm{X} | S}(\bm{x} | s = 0) \neq F_{\bm{X} | S}(\bm{x} | s = 1)$ under the assumptions. This is an important, even if simple, insight given that a researcher's first instinct may be to subset along the moderator variable. Similarly, as will be described later, common regression specifications that include interactions also generally fail to recover unbiased and consistent estimates of the ATME.

\subsection{Alternative Identification Strategies}

The parallel estimation framework also provides the opportunity to employ alternative within-subset identification strategies in place of selection on observables. Because of the randomization of the treatment, the data can be split into treatment-level subsets, and a causal effect of the moderator on the outcome within subset can be identified and estimated according to other identification strategies, provided that the relevant variables and additional identification features are pre-treatment. 

Specifically, given Assumptions \ref{assump:sutva} and \ref{assump:exog}, identification of the ATME could be achieved via an alternative strategy if (a) the additional requisite assumptions are met such that the strategy achieves identification of the causal effect of $S$ on $Y$ in both subsets $T_i = 0$ and $T_i = 1$, and (b) the auxiliary variables and features used in the alternative identification strategy are pre-treatment. If a regression discontinuity (or encouragement/instrumental variables) design were used as the within-subset identification strategy, the forcing variable (or encouragement/instrument and principal strata) would need to be pre-treatment or not otherwise impacted by the treatment. Using such identification strategies would also require viewing the ATME as an appropriately local estimand.

As an example, consider the encouragement/instrumental variables approach. Continue to make Assumptions \ref{assump:sutva} and \ref{assump:exog}, and now let $Z_i \in \{0,1\}$ denote for each unit $i$ the encouragement/assignment to the moderator $S_i \in \{0,1\}$. However, we allow for noncompliance with the encouragement/assignment such that for some units $Z_i \neq S_i$. In this approach, $Z_i$ replaces $\bm{X}_i$ as the auxiliary variable used in identification, and Assumption \ref{assump:pretre} is amended such that the moderator and instrument are pre-treatment. Then within each subset $T_i = 0,1$, a causal effect of $S$ on $Y$ can be identified using the local average treatment effect (LATE) framework applied to $Y$, $S$, and $Z$. In other words, $Z$ can be regarded as the instrument and $S$ as the treatment variable in the standard LATE framework, and the standard LATE assumptions then applied \citep{imbensangrist1994}. Then within each subset $T_i = 0,1$, it is possible to estimate (e.g. via the Wald IV estimator) the LATE, which is the average causal effect of $S$ on $Y$ for compliers, defined as units for whom it would always be true that $Z_i = S_i$ regardless of the value of $Z_i$ assigned. The difference between these two effects for each subset $T_i = 0,1$ then constitutes an estimate of the ATME for compliers.

\section{Implementing the Parallel Estimation Framework}
To provide concreteness, given a randomized treatment and non-randomized moderator and under Assumptions \ref{assump:sutva} -- \ref{assump:cs}, estimation of the ATME using the parallel estimation framework proceeds as follows:
\begin{enumerate}
\item Subset the data by \emph{treatment} level.
\item Separately for each treatment subset $T_i=0,1$, estimate the effect of $S$ on $Y$.
	\begin{itemize} \addtolength{\baselineskip}{-0.3\baselineskip}
	\item Use preferred method(s) of covariate adjustment (e.g. regression, matching, propensity score weighting, etc.) to perform the estimation controlling for $\bm{X}$.
	\item \emph{Alternative identification strategies (e.g. regression discontinuity, instrumental variables, selection on unobservables) could also be employed under alternative assumptions pertaining to the relationship between $S$ and $Y$, provided that any auxiliary variables used for identification are pre-treatment.}
	\end{itemize}
\item Calculate the difference between the two estimates from Step 2, which provides an estimate of the ATME, $\hat{\delta}$.
	\begin{itemize} \addtolength{\baselineskip}{-0.3\baselineskip}
	\item Provided that unbiased and/or consistent estimators of the effect of $S$ on $Y$ are employed in Step 2, then $\hat{\delta}$ is also unbiased and/or consistent. 
	\item $Var(\hat{\delta})$ can be approximated and estimated directly as the sum of the variances of the two estimates recovered in Step 2.
	\end{itemize}
\end{enumerate}

The following subsections will describe a number of specific implementations of the parallel estimation framework using covariate adjustment and illustrate the problems with using standard approaches to estimate the ATME.

\subsection{Parallel Within-Treatment Regressions}
The simplest method of estimation would be to use a set of parallel linear least squares regressions. In each regression, the outcome $Y$ is regressed on the moderator $S$, along with the full set of covariates $\bm{X}$ conditional upon which it is assumed that $S$ is independent of the potential outcomes. This yields estimates of the average causal effect of $S$ on $Y$ within each treatment level, and their difference constitutes the estimate of the ATME, $\hat{\delta}_{PR}$.
\begin{eqnarray} \nonumber 
Y_i &=& \hat{\alpha}_0 + \hat{\gamma}_0 S_i + \bm{X}_i' \hat{\beta}_0 + \hat{\epsilon}_i \:\:\:\:\:\:\:\:\:\:\: \forall  \:\:\:\:\:\: i : T_i = 0 \\ \nonumber
Y_j &=& \hat{\alpha}_1 + \hat{\gamma}_1 S_j + \bm{X}_j' \hat{\beta}_1 + \hat{\epsilon}_j \:\:\:\:\:\:\:\:\:\: \forall  \:\:\:\:\:\: j : T_j = 1 \\ \nonumber
\hat{\delta}_{PR} &=& \hat{\gamma}_1 - \hat{\gamma}_0 \\ \nonumber
Var(\hat{\delta}_{PR}) &\approx& Var(\hat{\gamma}_0) + Var(\hat{\gamma}_1)
\end{eqnarray}

\subsection{Full Interaction Regression}
The parallel regression equations also nest within a single regression equation. To see this, consider that the parallel regression equations can be combined as follows:
\begin{eqnarray}\nonumber
Y_i &=& \hat{\alpha}_0 + (\hat{\alpha}_1 - \hat{\alpha}_0)T_i + \hat{\gamma}_0 S_i + \bm{X}_i' \hat{\beta}_0 + (\hat{\gamma}_1 - \hat{\gamma}_0) T_i S_i + T_i \bm{X}_i' (\hat{\beta}_1 - \hat{\beta}_0) + \hat{\epsilon}_i \\
 &=& \hat{\alpha} + \hat{\tau} T_i + \hat{\omega} S_i + \bm{X}_i' \hat{\beta} + \hat{\delta}_{FR} \: T_i S_i + T_i \bm{X}_i' \hat{\xi} + \hat{\epsilon}_i \label{eq:singlereg}
\end{eqnarray}
As a result, $\hat{\delta}_{FR} = \hat{\delta}_{PR}$; the estimates are equivalent using both the parallel regression framework and the full interaction regression as specified above.

Note that $ T_i \bm{X}_i'$ corresponds to the interactions between the treatment and all of the control variables. This highlights the problem---a problem that is common among researchers seeking to estimate moderation effects---with estimating an ATME in a single regression framework by merely controlling for the covariates by themselves and only including the interaction between the treatment and moderator. As is made clear by equation (\ref{eq:singlereg}), such a specification would not recover an unbiased or consistent estimate of the ATME. As a simple case, consider a data-generating process determined by the following model:
$$Y_i = \alpha + \tau T_i + \omega S_i + \beta X_i + \delta T_i S_i + \xi T_i X_i + \epsilon_i$$
where $\epsilon$ is distributed with mean zero and, for simplicity, $X$ denotes a single covariate. This model corresponds to a situation where $X$ not only affects the outcome itself but also, similar to the moderator of interest $S$, interacts with the treatment's effect on the outcome. Hence, it is not sufficient to simply include $X$ as a regression control variable by itself in order to estimate the ATME, as omission of the $T  X$ term would lead to omitted variable bias that affects $\hat{\delta}_{FR}$ given a non-zero value of $\xi$ and a non-zero covariance between $S$ and $X$. 

An interesting special feature to note about this linear context, already evident from the discussion above, is that the full interaction regression simultaneously recovers an estimate of the causal moderation effect of $S$ (i.e. the ATME) as well as estimates of the causal moderation effects of any number of other variables contained in $\bm{X} = (X_1, ..., X_k , ...)'$ in equation (\ref{eq:singlereg}), provided that those variables also satisfy the necessary conditions. In particular, each must also be manipulable at the individual level to have a well-defined causal moderation effect, and each must be independent of the potential outcomes conditional on the other included variables. If either of these conditions is not met for any given variable $X_k$, then while inclusion of $\xi_k T X_k$ in the equation (\ref{eq:singlereg}) specification is necessary for estimation of the ATME, $\xi_k$ itself does not correspond to an identified causal moderation effect. Note, however, that the possibility of estimating multiple causal moderation effects simultaneously does not generally carry over to other methods of estimation researchers would likely be interested in employing, as presented below.

\subsection{Parallel Matching and Alternative Methods}

Linear regression approaches to estimating causal effects, whether for the ATME as described above or for individual ATEs with observational data, rely upon specific model assumptions. This includes, of course, linearity. In addition, less obvious is the implicit assumption of constant effects given that linear regression in general performs a conditional-variance-weighting scheme that diverges from the type of weighting necessary to achieve proper average effect estimates \citep[see][chapter 3]{angristpischke2009}. These assumptions may be undesirable given certain data and when the estimation results are likely to be sensitive to functional form and model specifications in general. Thus, researchers may wish to avoid the model dependence of linear regression (or any other parametric method for that matter) in the parallel estimation process. 

Hence, an attractive nonparametric alternative for estimating the ATME would be to use parallel within-treatment-subset matching methods. The parallel matching process is analogous to the parallel regression approach:
\vspace{0.0cm}
\begin{enumerate} \addtolength{\baselineskip}{-0.4\baselineskip}
\item Split sample into treated and control subsets. 
\item Within each subset, match ``moderated" units ($S_i=1$) and ``unmoderated" units ($S_i=0$) on the confounding variables ($\bm{X}_i$). Within each matched sub-sample, estimate the effect of $S$ on $Y$.
\item Compute the difference between these two estimated effects, which provides an estimate of the ATME.
\end{enumerate}
As with the implementation of matching in the ATE context, standard diagnostics can be performed for both the treated and control sub-samples to assess common support and the quality of the matching process in producing covariate balance, and various post-matching techniques can be applied to estimate the within-subset effects of $S$ on $Y$ \citep{ho2007matching, stuart2010matching}. More generally, any method of causal effect estimation under selection on observables can be applied to step 2 above. See Appendix B for additional details.

\subsection{Comparing the Parallel Estimation Framework to Conventional Approaches}

The two standard approaches most commonly used in the literature to estimate moderation effects are subgroup analysis and interaction regressions. As the results presented in this study demonstrate, these conventional methods generally do not constitute unbiased or consistent estimators of the ATME given a non-randomized moderator. The first approach, subgroup analysis, involves subsetting the data by different levels of the moderator variable, estimating the treatment effect within subsets, and calculating the difference between the two. This may be referred to alternatively as treatment-by-covariate interactions and comparison of conditional average treatment effects (CATEs) \citep[cf.][]{gerbergreen2012}. Except in the special case where both the treatment and moderator are randomly assigned, such estimates can clearly not be interpreted as causal moderation effects.

The second common approach is using a controlled interaction regression, in which the outcome is regressed on the treatment, the moderator, the treatment-moderator interaction, and control variables. As already shown above, this approach also falls short for estimating causal moderation in the case of a randomized treatment and non-randomized moderator. While this result has already been observed in specific fields of scholarship, including social psychology \citep{yzerbytetal2004, hulletal1992} and gene-environment interaction \citep{keller2014gene}, it appears not to be widely appreciated. Instead, empirical researchers continue to investigate causal moderation phenomena using interaction regressions that control for possible confounders but not their interactions with the treatment.

The controlled interaction regression specification can, of course, be extended for estimation of causal moderation by including additional interaction terms. That this appears to be done rarely suggests that, in spite of the large corpus of methodological literature on interaction terms in regression \cite[e.g.][]{aikenwest1991, bramboretal2006, hainmueller2019much, kamfranzese2009}, there remains a lack of clarity on how to choose which interaction terms to include in one's regression model and which to omit. One problem is that interaction/moderation effects are often discussed without explicitly stating what type of estimand (descriptive versus causal) is of interest. For more discussion on this, see \citet[][pp. 268--270]{vanderweele2015} and \citet{kraemeretal2008}. A second problem is that interaction/moderation effects are often discussed without reference to whether the treatment and moderator of interest are exogenous or randomized. The result is a lack of explicitness in terms of what types of confounding are possible and, accordingly, what types of adjustments must be made in response.

In sum, while there already exists extensive methodological literature on interaction and moderation effects, the disparate assumptions and analytical goals tacitly underlying different recommendations makes it difficult for researchers to know exactly what to do given their own specific data and goals. Furthermore, because most of the literature is anchored within the regression context, the recommendations often lack general formal results to justify certain specifications; researchers must instead fall back on the usual, slightly uncomfortable assumption that the posited parametric model is correct. This is not reassuring for researchers who want a robust model that is both motivated by clear-cut theoretical guidance yet also acknowledges that the estimation models we specify are rarely correct or complete. By formalizing the definition of a causal moderation effect in the form of the ATME and discussing it within the specific design context of a randomized treatment and non-randomized moderator, this study clarifies for experimentalists precisely what is meant by causal moderation, what assumptions its identification requires, and how its estimation can be undertaken.

\section{Applications}

This section presents two applications to provide concrete illustrations of the parallel estimation framework. Both applications come from previous studies involving experimentally manipulated treatments, and under consideration is the extent to which a separate, non-randomized variable moderates the treatment effect. Thus, the applications fall within the present scope of dealing with a randomized treatment and non-randomized moderator. Furthermore, both applications deal with moderators that are mutable at the individual level.

As already described, in addition to randomization of the treatment, identification of the ATME requires a number of other assumptions. Of particular note is the observation of a set of control variables ($\bm{X}$) conditional upon which the moderator is independent of the potential outcomes. Each application specifies a plausible set to meet this condition. Yet just as in the case of estimating ATEs for non-randomized treatments, this conditional independence assumption is non-refutable, and subject matter experts may identify other variables missing from the conditioning set. Note, however, that the applications are meant to be illustrative, and they provide the structure for further research should the correct conditioning set be measured in the future. In addition, identification of the ATME also requires the SUTVA assumption. In both applications below, the experimental designs make SUTVA defensible within the study contexts. However, this of course does not guarantee that SUTVA would be met in less highly controlled contexts, or that more real-world versions of the treatment would have similar effects. Finally, in both applications, the samples were not drawn through pure simple random sampling. However, we may adopt the sample-as-population approach and interpret the estimates as pertaining to sample estimands \citep{imai2008misunderstandings}.

\subsection{Framing Experiment on Support for Public Assistance}

Many scholars of American voter behavior and political psychology study how issue frames can be used by politicians to influence public attitudes, either positively or negatively, toward specific policies. One of the simplest yet most powerful and prominent examples is the framing of policies on public assistance to the poor as ``welfare." While the term welfare is simply another label for public assistance, scholars have long noted that substantial segments of the U.S. voting population hold a peculiar aversion toward this label \citep{kluegelsmith1986,smith1987,rasinski1989,shawshapiro2002}. Furthermore, years of concrete empirical evidence of this phenomenon have been compiled, after the General Social Survey (GSS) beginning in the 1980s introduced a question-wording experiment that measures the effect of using the label ``welfare" rather than ``assistance to the poor" on support for assistance to the poor. As the results across years of GSS waves have shown, referring to welfare causes enormous increases---on the order of 30-50 percentage points---in the probability of a respondent asserting that ``too much" is being spent on such policies in the United States.

The experimental evidence in the GSS has led to an abundance of subsequent scholarship focused on explaining the underlying sources of this ``welfare-label effect" and variation in its magnitude across different subsets of voters. In one example, \cite{greenkern2012} develop a method of applying Bayesian Additive Regression Trees (BART) to model and discover treatment effect heterogeneity and apply the method to the GSS welfare framing experiment data. In their study, they find that substantial amounts of heterogeneity in the welfare-label effect manifest as a function of a number of covariates, including party identification, with the welfare-label effect being more pronounced among Republicans than Democrats.

However, Green and Kern's finding that the welfare-label effect magnitude depends upon party identification does not, nor do Green and Kern claim it to, demonstrate that party identification itself is causing variation in the welfare effect. There are, of course, reasons to believe that party identification itself may moderate the welfare effect to some extent. A voter's party identification relates directly to how responsive that voter is to party cues, and party cues involve delivery of partisan ideological messages, position preferences, and issue frames. Given the economically conservative preference for limited redistribution of the Republican Party, it is reasonable to expect that Republican Party cues to voters include calls for limiting social assistance, and that such calls might frame social assistance using the stigmatized label of welfare. For these reasons, we might expect that identification with the Republican Party would, in fact, make an individual more sensitive to the welfare frame and hence exhibit a stronger welfare-label effect than had s/he not adopted that party identification. Furthermore, even if most voters do not change their party identification often or at all, party identification is a characteristic that is clearly mutable at the individual level.

In contrast, however, there are many non-political factors known to influence voters' party identification choices that may also make voters more sensitive to the welfare frame for various reasons not directly related to party identification itself. For instance, some researchers have explored the possibility of interactions between the welfare-label effect and such variables as education and racial perceptions \citep{federico2004welfare, henry2004hate}. Such variables may also correlate with party identification such that any conditionality in the welfare-label effect across voters with different party identifications may be at least partially the result of these other variables. 

To investigate the causal moderation of the welfare effect attributable to party identification, the parallel estimation framework is applied to the GSS data, and estimates of the ATME using parallel regression and parallel matching are compared to moderation effect estimates using conventional methods. As described above, the randomized treatment is whether the respondent was asked about ``welfare" (coded as 1) or ``assistance to the poor" (coded as 0), and the outcome is whether the respondent stated that ``too much" is being spent on such policies (coded as 1) or that ``too little"/``about the right amount" is being spent on such policies (coded as 0). The moderator is a dichotomous version of party identification, with those respondents identifying as Republicans coded as 1 and all others coded as 0. Covariates used as control variables include age, sex, race, highest education level, real income, number of children, region, city/town type (based on size and urbanicity), year surveyed, and racial attitudes. While scholars of American voting behavior have long considered racial attitudes to be possible determinants of party identification \citep{carminesstimson1989}, some studies have questioned the precise relationship between racial attitudes and party identification \citep[e.g.][]{abramowitz1994}. Though the two are likely co-determined, for the purposes of the present investigation, it is illuminating to isolate the party identification moderation effect that is not related to racial attitudes. For this reason, racial attitudes are included as a control variable, based on the coding used by \cite{greenkern2012}.

\begin{figure}[!htb]
\centering
\caption{Moderation of Welfare Framing Effect by Partisanship}\label{fig:welfmod}
\includegraphics[scale=0.7]{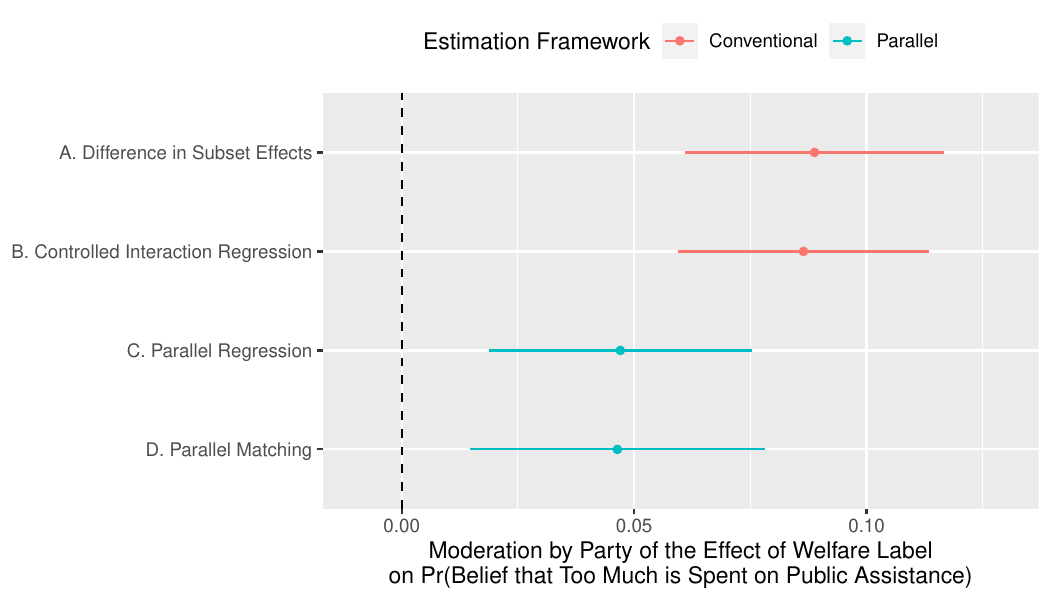}
\end{figure}

Figure \ref{fig:welfmod} presents the results, with point estimates and 95\% confidence intervals estimated using the conventional approaches and parallel estimation framework respectively.
As illustrated in the upper half of the plot, the results from the conventional estimators suggest substantial moderation. The first estimator is the difference in subset effects, where the two subsets are Republican respondents and non-Republican respondents, and each subset effect is the simple difference between the probability of believing that too much is being spent on poverty assistance in the treatment condition (i.e. given the ``welfare" frame) and the probability of believing that too much is being spent on poverty assistance in the control condition (i.e. given the ``assistance to the poor" frame). The difference in subset effects highlights important heterogeneity of the welfare effect: while there is a welfare effect among all partisan groups, the magnitude of the estimated welfare effect is almost 10 percentage points higher among Republicans than among other respondents. Specifically, the estimated effect among non-Republicans is 0.329, while it grows over 25\% to 0.418 for Republicans.

The second estimator whose results are presented in Figure \ref{fig:welfmod} is the controlled interaction regression estimator, which is the estimate of the interaction effect between the treatment and moderator in a linear regression of the outcome on the treatment, moderator, treatment-moderator interaction, and control variables. While a number of the control variables are correlated with party identification (as should be expected), and while we should also expect a number of these control variables to bear upon voters' susceptibility to the welfare effect, the figure shows that the inclusion of controls in the interaction regression has virtually no effect on the moderation effect estimate, compared to the simple difference in subset effects.

As explained in this study, it would be incorrect to interpret the upper two estimates in Figure \ref{fig:welfmod} as causal moderation effects because they are based on estimators that do not perform covariate adjustment in a way that allows for estimation of an identified causal moderation estimand. In contrast, ATME estimates using the parallel regression method and the parallel matching method are reported in the lower half of Figure \ref{fig:welfmod}. The matching procedure---implemented with the \texttt{Matching} package in \texttt{R} \citep{JSSv042i07}---used Mahalanobis distance and, within each \emph{treatment-level subset}, matched each Republican (non-Republican) unit with the closest non-Republican (Republican) unit with replacement. As can be seen, these results paint a much more conservative picture of the extent to which party identification \emph{per se} moderates the welfare effect. Specifically, the estimates yielded by these methods are attenuated toward zero and cut roughly in half. 

These results indicate that a substantial portion of the moderation effect attributed to party identification by the conventional estimators is actually the result of party identification being correlated with other variables that themselves moderate the welfare effect. Indeed, the results from the parallel regression method include statistically significant interactions between the treatment and several other variables---including education level, real income, city/town type, and racial attitudes---that are also correlated with party identification.

\subsection{Asylum Seeker Conjoint Experiment}

European governments have recently struggled with the largest refugee crisis since World War II, with Europe receiving approximately 1.3 million new asylum claims in 2015. Exacerbating the political difficulties of accommodating asylum seekers while minimizing domestic public backlash is that, while most asylum seekers currently originate from Muslim-majority countries, research has shown that European voters would prefer to grant asylum to non-Muslim applicants \citep{bansaketal2016}. Furthermore, some right-wing parties have sought to mobilize citizens around asylum policy issues, appealing to nationalistic impulses and promoting anti-asylum policies in order to increase their vote shares \citep{dustmann2018refugee, dinas2019waking, hangartner2019does}. This episode is important in and of itself, and it also represents an example of a longstanding theme explored by political scientists whereby nationalism is considered a malleable dimension of public opinion that may fluctuate over time, is endogenous to current events, and constitutes a potential target for manipulation via government intervention and elite political communication \citep[e.g.][]{mueller1970presidential, baker2001patriotism, weiss2013authoritarian}. In this case, we may ask: to what extent might nationalism increase Islamophobia? More specifically, does increased nationalism strengthen voters' preferences for non-Muslim asylum seekers?

To investigate, the parallel estimation framework is applied to data from a previous conjoint experiment embedded in an online survey fielded in fifteen European countries with approximately 18,000 respondents \citep{bansaketal2016}. In the conjoint experiment, respondents were presented with pairs of hypothetical asylum-seeker profiles comprised of various attributes with randomly assigned levels. The analysis here focuses on one attribute in particular, which serves as the randomly assigned treatment $T$ in the parallel estimation process: whether the asylum seeker was Muslim (1) or Christian (0). For each pair of asylum seekers, the respondent was asked to specify which s/he preferred, providing a binary preference outcome variable $Y$. Each respondent's level of nationalism was also measured via a set of questions in the survey combined into an index. This measure is dichotomized into a binary measure of high-versus-low nationalism, with the median index value chosen as the cut point, for use as the moderator variable $S$. Thus, being investigated is whether nationalism moderates the effect of an asylum seeker's religion on voters' acceptance of the asylum seeker. Implicit in this investigation is the assumption that nationalism is indeed mutable at the individual level, in line with previous political science scholarship that treats nationalism as endogenous to social and political factors. Finally, a range of demographic and other characteristics are used as the set of control variables $\bm{X}$. This includes gender, age, income, education, and left-right political ideology. More details on the study design and variables can be found in Appendix C.

Figures \ref{fig:convest} and \ref{fig:parest} present the results, pooled and across countries, with moderation effect point estimates and 95\% confidence intervals estimated using the conventional approaches and parallel estimation framework respectively.
\begin{figure}[!htb]
    \centering
    \begin{minipage}{.5\textwidth}
        \centering
        \caption{Conventional Estimators}\label{fig:convest}
        \includegraphics[scale=0.43]{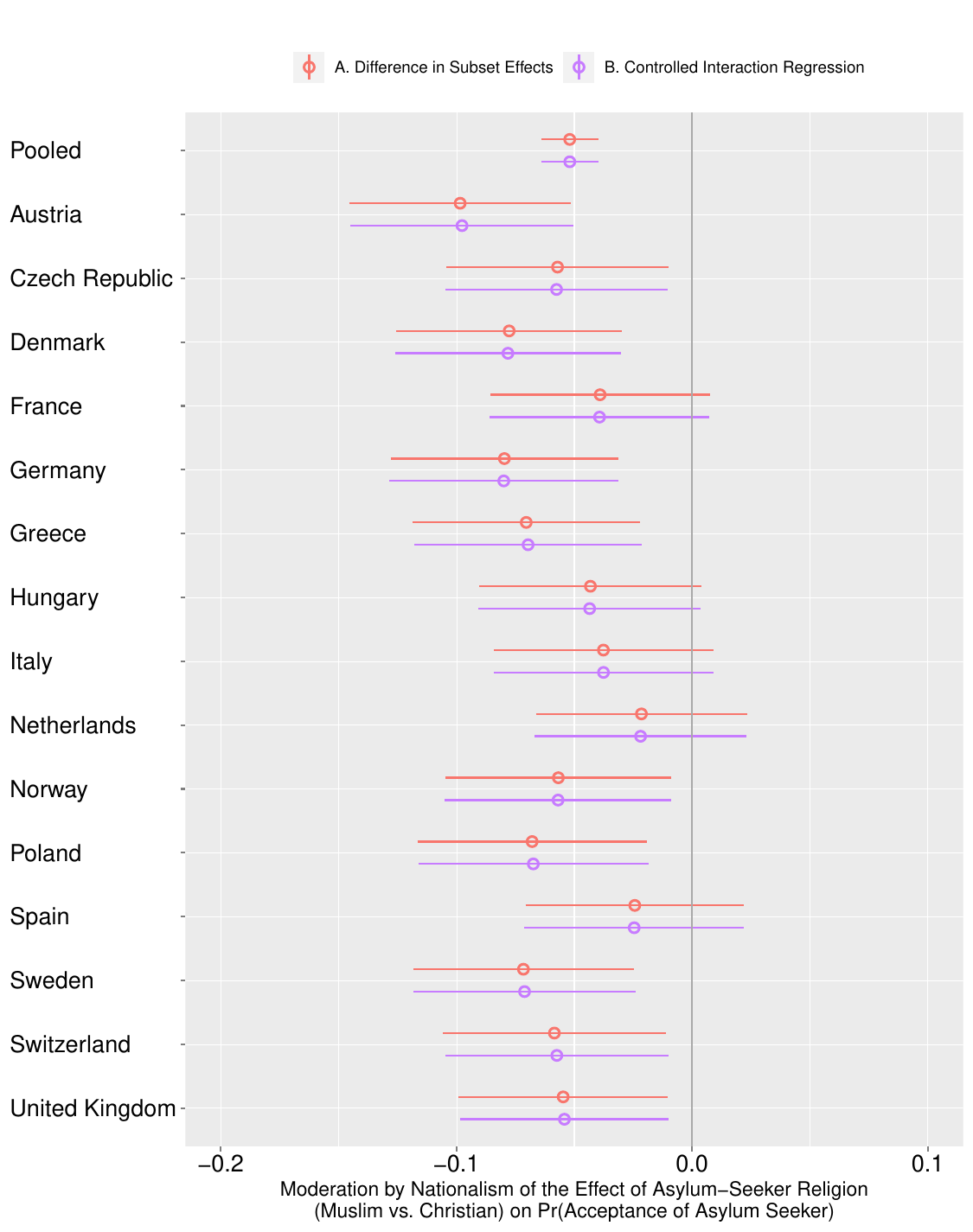}
    \end{minipage}%
    \begin{minipage}{0.5\textwidth}
        \centering
        \caption{Parallel Framework Estimators}\label{fig:parest}
        \includegraphics[scale=0.43]{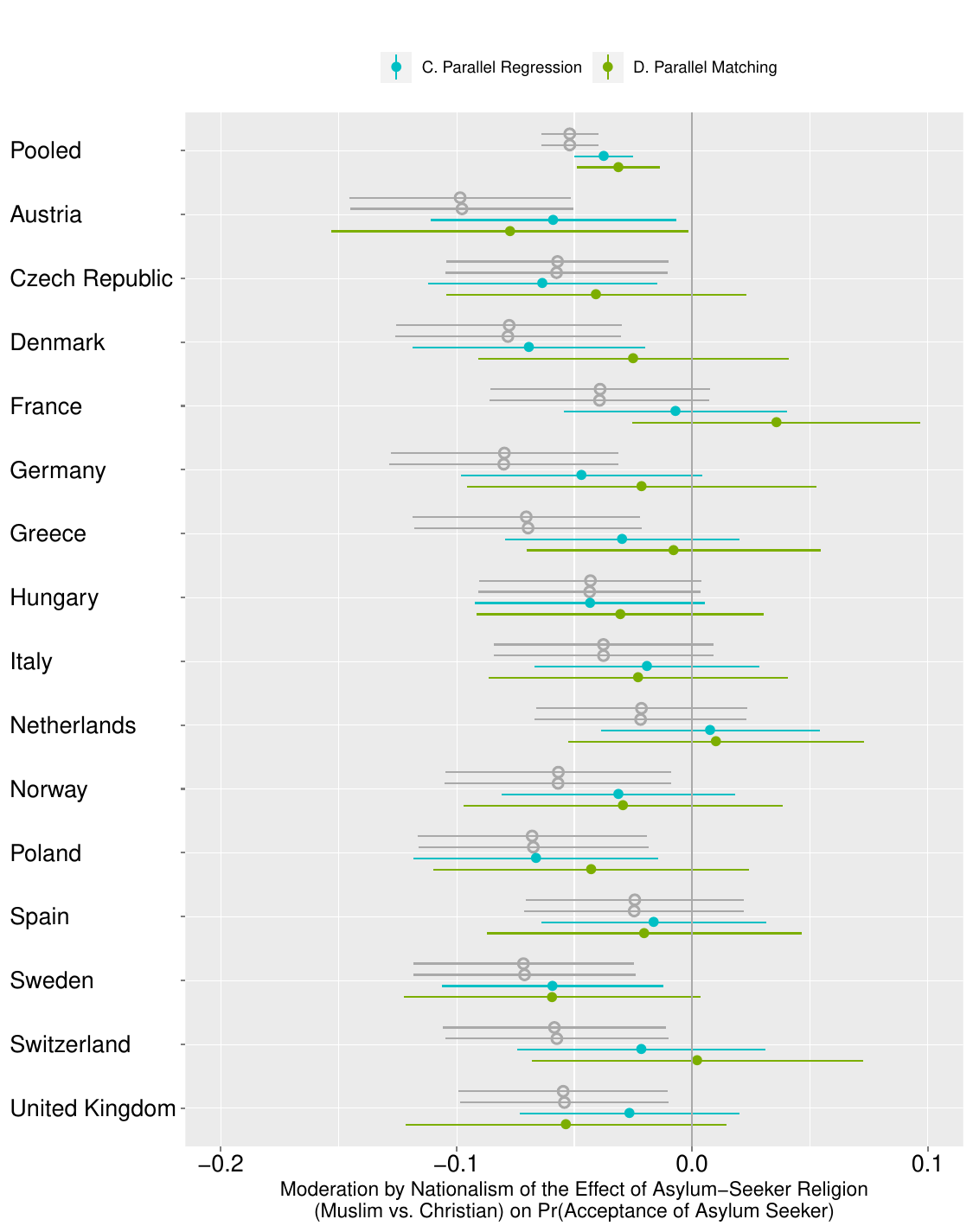}
    \end{minipage}
\end{figure}
The results from the conventional estimators, shown in Figure \ref{fig:convest}, highlight strong heterogeneity of voters' preferences regarding the religion of asylum seekers: the lower probability of accepting a Muslim asylum seeker relative to a Christian asylum seeker is more pronounced among high-nationalism respondents than low-nationalism respondents in all countries, a statistically significant finding at $\alpha = 0.05$ in ten out of the fifteen countries included in the survey. Substantively, the magnitude of that heterogeneity varies by country. For instance, in Austria, the difference in subset effects is about $-0.1$, meaning that the negative gap in the probability of accepting a Muslim asylum seeker relative to a Christian asylum seeker grows by 10 additional percentage points when moving from the low-nationalism segment of voters to the high-nationalism segment.

In contrast, however, the causal moderation results using the parallel estimation framework, shown in Figure \ref{fig:parest}, paint a more conservative picture of the extent to which increased nationalism can strengthen Islamophobic biases toward asylum seekers. Specifically, with only a few exceptions the causal moderation effect estimates yielded by these methods are attenuated toward zero compared to the conventional estimates. Of the ten countries where the conventional estimators yielded statistically significant moderation effects, only one is found to have statistically significant causal moderation effects according to both the parallel regression and parallel matching estimators, and only four others are found to have statistically significant causal moderation effects by either estimator. Note that the matching procedure used Mahalanobis distance and matched all of the moderated (high nationalism) units with the closest unmoderated (low nationalism) units, with replacement. Hence the matching estimates should be interpreted as the ATMEs for the highly nationalistic.

\section{Discussion and Conclusions}

\subsection{Sensitivity Analysis}

Identification of the ATME via selection on observables relies on the assumption that all variables that confound the relationship between the moderator and the outcome are conditioned upon. Of course, as with the estimation of ATEs using observational data, it is impossible to know and difficult to argue that every possible confounder has been observed. Instead, the practical question is: How serious is the bias likely to be due to unobserved confounders, and how sensitive is the estimate of interest---the ATE in the simple context, and the ATME in the context of this study---to the existence of such remaining bias? Answering this question can be done by pairing subject-matter expertise with a rigorous sensitivity analysis. The parallel estimation framework allows for an adaptation of ATE sensitivity analysis to the causal moderation context. Appendix D in the SM develops and presents an ATME sensitivity analysis, along with an application.

\subsection{Scope Conditions}

As already discussed, for the ATME as conceptualized in this study to be a relevant and fundamentally meaningful estimand of interest, it must be the case that the moderator in question is indeed mutable at the individual level (just as a treatment variable must be mutable for a clearly defined ATE). Thus, for researchers interested in investigating causal moderation and estimating an ATME for a particular moderator, they must first ask themselves whether such a goal makes sense given the nature of the moderator variable. Moderators that are immutable characteristics, such as the research subjects' racial identities, may fall out of the scope of causal moderation and should instead be investigated within the estimation and interpretive frameworks of treatment effect heterogeneity (though note that in some studies it is not racial identity but rather perception of racial identity that is of interest, and the latter can be manipulated). Also note that, as presented in this study, causal moderation involves a moderator that is itself unaffected by the treatment of interest. For this reason, it is a separate phenomenon from causal \emph{mediation}, and the ATME is distinct from estimands formulated within the causal mediation context.

Finally, consider the case in which the treatment is not randomly assigned but believed to adhere jointly to the same selection on observables as the moderator. In other words, in place of assumptions \ref{assump:exog} -- \ref{assump:soo}, we would assume $(Y_i(1,1), Y_i(0,1), Y_i(1,0), Y_i(0,0)) \indep (T_i, S_i) \: \big| \: \bm{X}_i$. Under these conditions, identification of the ATME would still hold \citep[see][for a general treatment of this case]{vanderweele2009distinction, vanderweele2015}. However, the estimation process could not be parallelized into two separate components in the manner presented in this study because the joint distribution of the moderator and control variables would not be equivalent across subsets $T_i=0,1$. As described, the attraction of the parallel estimation framework is in the additional clarity and accessibility it provides to applied researchers, as well as the tractability it gives for nonparametric procedures like matching. However, its drawback is that in practice it requires a randomized treatment.

\subsection{Conclusions}

If a researcher's goal is to estimate causal moderation effects, it is generally preferable to design a factorial experiment in which the treatment and moderator are both randomized. Unfortunately, however, randomization of the moderator is often not possible, in which case identification of causal moderation effects requires additional work in the analysis stage. For researchers who find themselves in such a situation, this study has introduced a straightforward, generalized framework for estimating causal moderation effects.

This study has sought to present important insights and useful tools for experimental researchers. First, by employing the potential outcomes model to explicitly define causal moderation, distinguishing it from treatment effect heterogeneity, and laying out a targeted set of assumptions for its identification, the study highlights formally how conventional estimation approaches used by experimentalists do not constitute consistent or unbiased estimators of causal moderation effects. Second, it offers researchers a simple and flexible approach for the estimation of causal moderation effects, allowing researchers to implement their preferred method of covariate adjustment, including both parametric and non-parametric methods, or alternative identification strategies of their choosing. In addition, the framework also provides a set-up whereby sensitivity analysis designed for the ATE context can be extended to the causal moderation context.

Ultimately, the value of investigating causal moderation is in allowing researchers to better understand and anticipate the efficacy of a particular treatment should other variables (i.e. moderators) change or be intervened upon when the treatment is administered or investigated in other contexts and/or in the future. This endeavor is core to the broader, critical, but often elusive goal of understanding the generalizability of treatment effects.

\setstretch{1.00}
\setlength{\bibsep}{8pt}

\renewcommand{\bibsection}{\section{References}}

\bibliographystyle{apalike}
\bibliography{refs}

\clearpage

\renewcommand{\thepage}{\roman{page}}
\setcounter{page}{0}

\Large
\begin{center}
\textbf{Supplementary Materials} \\ 
\vspace{1cm}
for \\
\vspace{4cm}
\LARGE
Estimating Causal Moderation Effects with Randomized Treatments and Non-Randomized Moderators \\
\vspace{1cm}
\Large
Kirk Bansak
\end{center}

\normalsize

\clearpage

\addtolength{\baselineskip}{0.1\baselineskip}

\setcounter{section}{0}
\setcounter{subsection}{0}
\renewcommand{\thesection}{\Roman{section}}
\renewcommand{\thesubsection}{\Alph{subsection}}

\makeatletter
\renewcommand{\section}{\@startsection{section}{1}{0mm}{-\baselineskip}{0.25\baselineskip}{\centering\normalfont\normalsize\scshape}}

\section*{Appendix A: Proofs}

\subsection*{Proof of Proposition 1}

\bigskip

To begin, note that Assumptions \ref{assump:pretre} and \ref{assump:cs} together ensure joint common support: 
$$0 < P(T_i = t,S_i = s|\bm{X}_i) < 1$$ 
for any $s=0,1$ and $t=0,1$. This follows from Assumption \ref{assump:cs} that $0 < P(S_i = s|\bm{X}_i) < 1$, and Assumption \ref{assump:pretre} implying that $T_i \indep (S_i, \bm{X}_i)$ and hence that $0 < P(T_i = t|\bm{X}_i) < 1$ and $P(T_i = t,S_i = s|\bm{X}_i) = P(T_i = t|\bm{X}_i)P(S_i = s|\bm{X}_i)$. Thus, $0 < P(T_i = t,S_i = s|\bm{X}_i) < 1$.

\bigskip

Joint common support then allows for defining $E[Y_i | T_i = t, S_i = s, \bm{X}_i]$ for all $t=0,1$ and $s=0,1$, which thereby allows for defining:

$$\{ E[Y_i|T_i=1,S_i=1,\bm{X}_i] - E[Y_i|T_i=0,S_i=1,\bm{X}_i] \}$$
$$ - \{ E[Y_i|T_i=1,S_i=0,\bm{X}_i] - E[Y_i|T_i=0,S_i=0,\bm{X}_i] \}$$
$$= E[Y_i|T_i=1,S_i=1,\bm{X}_i] - E[Y_i|T_i=0,S_i=1,\bm{X}_i]$$
$$ - E[Y_i|T_i=1,S_i=0,\bm{X}_i] + E[Y_i|T_i=0,S_i=0,\bm{X}_i]$$
Given $Y_i = Y_i(T_i,S_i)$, this equals:
$$E[Y_i(1,1)|T_i=1,S_i=1,\bm{X}_i] - E[Y_i(0,1)|T_i=0,S_i=1,\bm{X}_i]$$ 
$$- E[Y_i(1,0)|T_i=1,S_i=0,\bm{X}_i] + E[Y_i(0,0)|T_i=0,S_i=0,\bm{X}_i]$$

\bigskip

Now, note that Assumption \ref{assump:pretre} means that for any $t=0,1$ and $s=0,1$, the conditional distribution of $Y_i(t,s)$ has the following properties:
$$f(Y_i(t,s)|T_i=t,S_i=s,\bm{X}_i) = \frac{f(Y_i(t,s),T_i=t,S_i=s,\bm{X}_i)}{f(T_i=t,S_i=s,\bm{X}_i)} = \frac{f(Y_i(t,s),S_i=s,\bm{X}_i)P(T_i=t)}{f(S_i=s,\bm{X}_i)P(T_i=t)}$$
$$= \frac{f(Y_i(t,s),S_i=s,\bm{X}_i)}{f(S_i=s,\bm{X}_i)} = f(Y_i(t,s)|S_i=s,\bm{X}_i)$$
and hence
$$E[Y_i(t,s)|T_i=t,S_i=s,\bm{X}_i] = E[Y_i(t,s)|S_i=s,\bm{X}_i]$$

\bigskip

Further, note that for any $s=0,1$, given Assumption \ref{assump:soo},
$$E[Y_i(t,s)|S_i=s,\bm{X}_i] = E[Y_i(t,s)|\bm{X}_i]$$

\bigskip

All of this implies that
$$E[Y_i(1,1)|T_i=1,S_i=1,\bm{X}_i] - E[Y_i(0,1)|T_i=0,S_i=1,\bm{X}_i]$$ 
$$- E[Y_i(1,0)|T_i=1,S_i=0,\bm{X}_i] + E[Y_i(0,0)|T_i=0,S_i=0,\bm{X}_i]$$

$$= E[Y_i(1,1)|\bm{X}_i] - E[Y_i(0,1)|\bm{X}_i] - E[Y_i(1,0)|\bm{X}_i] + E[Y_i(0,0)|\bm{X}_i]$$

\bigskip

Hence,
$$\delta = $$
$$E[Y_i(1,1)] - E[Y_i(0,1)] - E[Y_i(1,0)] + E[Y_i(0,0)]$$
$$= E_{\bm{X}} \Big[ \:\: E[Y_i(1,1)|\bm{X}_i] - E[Y_i(0,1)|\bm{X}_i] - E[Y_i(1,0)|\bm{X}_i] + E[Y_i(0,0)|\bm{X}_i] \:\: \Big]$$
$$= E_{\bm{X}} \Big[ \:\:  E[Y_i|T_i=1,S_i=1,\bm{X}_i] - E[Y_i|T_i=0,S_i=1,\bm{X}_i]  $$ 
$$  -  E[Y_i|T_i=1,S_i=0,\bm{X}_i] + E[Y_i|T_i=0,S_i=0,\bm{X}_i]  \:\: \Big]  $$
$\square$

\clearpage

\subsection*{Proof of Proposition 2 (See Appendix B)}

\bigskip

$$E[Y_i|T_i=1,S_i=1,\bm{X}_i] - E[Y_i|T_i=1,S_i=0,\bm{X}_i] + E[Y_i|T_i=0,S_i=0,\bm{X}_i] - E[Y_i|T_i=0,S_i=1,\bm{X}_i]$$
$$= E \left[ \frac{Y_i}{\pi_{11}(\bm{X}_i)} \Big| T_i=1,S_i=1,\bm{X}_i \right] \pi_{11}(\bm{X}_i) + E \left[\frac{-Y_i}{\pi_{10}(\bm{X}_i)} \Big| T_i=1,S_i=0,\bm{X}_i \right] \pi_{10}(\bm{X}_i)$$
$$+ E \left[\frac{Y_i}{\pi_{00}(\bm{X}_i)} \Big| T_i=0,S_i=0,\bm{X}_i \right] \pi_{00}(\bm{X}_i) + E \left[\frac{-Y_i}{\pi_{01}(\bm{X}_i)} \Big| T_i=0,S_i=1,\bm{X}_i \right] \pi_{01}(\bm{X}_i)$$
where $\pi_{ts}(\bm{X}_i)$ refers to the probability a unit receives $T_i=t$ and $S_i=s$ as a function of $\bm{X}_i$.

\bigskip

Given Assumptions \ref{assump:exog} and \ref{assump:pretre},
$$P(T_i=t, S_i=s |\bm{X}_i) = P(T_i=t)P(S_i=s |\bm{X}_i)$$

\smallskip

Let $P(T_i=1) = p \: , \:\: P(T_i=0) = 1-p \: , \:\: P(S_i=1|\bm{X}_i) = \pi(\bm{X}_i)  \: , \:\:P(S_i=0|\bm{X}_i) = 1- \pi(\bm{X}_i)$, \\
which simplifies the expression above to:
$$E \left[ \frac{Y_i}{p \pi(\bm{X}_i)} \Big| T_i=1,S_i=1,\bm{X}_i \right] p \pi(\bm{X}_i)$$
$$ + E \left[\frac{-Y_i}{p(1-\pi(\bm{X}_i))} \Big| T_i=1,S_i=0,\bm{X}_i \right] p(1-\pi(\bm{X}_i))$$
$$+ E \left[\frac{Y_i}{(1-p)(1-\pi(\bm{X}_i))} \Big| T_i=0,S_i=0,\bm{X}_i \right] (1-p)(1-\pi(\bm{X}_i))$$
$$ + E \left[\frac{-Y_i}{(1-p)\pi(\bm{X}_i)} \Big| T_i=0,S_i=1,\bm{X}_i \right] (1-p)\pi(\bm{X}_i)$$
which, using the conditional values and the law of total expectation, becomes: 
$$E \left[ Y_i\frac{(T_i-p)(S_i-\pi(\bm{X}_i))}{p(1-p)\pi(\bm{X}_i)(1-\pi(\bm{X}_i))} \Big| \bm{X}_i \right]$$

\smallskip

Taking the expectation, we achieve:
$$E_{\bm{X}} \Big[ E[Y_i|T_i=1,S_i=1,\bm{X}_i] - E[Y_i|T_i=1,S_i=0,\bm{X}_i]$$
$$+ E[Y_i|T_i=0,S_i=0,\bm{X}_i] - E[Y_i|T_i=0,S_i=1,\bm{X}_i] \Big]$$
$$ = \delta$$
$$=E_{\bm{X}} \left[ E \left[ Y_i\frac{(T_i-p)(S_i-\pi(\bm{X}_i))}{p(1-p)\pi(\bm{X}_i)(1-\pi(\bm{X}_i))} \Big| \bm{X}_i \right] \right]$$
$$=E \left[ Y_i\frac{(T_i-p)(S_i-\pi(\bm{X}_i))}{p(1-p)\pi(\bm{X}_i)(1-\pi(\bm{X}_i))} \right]$$
$\square$
\clearpage

\section*{Appendix B: Additional Details on Parallel Estimation Methods}

\setcounter{table}{0}
\setcounter{figure}{0}
\setcounter{subsection}{0}
\renewcommand{\thetable}{B\arabic{table}}%
\renewcommand{\thefigure}{B\arabic{figure}}%

\subsection{Propensity Score Weighting}

Propensity score weighting offers another method to apply within the parallel estimation framework to estimate the ATME. As with regression, this can be performed separately for each treatment-level subset, or the parallel paths can be combined into a single expression as follows:

\vspace{0.4cm}
\begin{proposition} \addtolength{\baselineskip}{-0.4\baselineskip} 
Given Assumptions \ref{assump:sutva}-\ref{assump:cs},
$$\delta =E \left[ Y_i \frac{(T_i -p)(S_i -\pi(\bm{X}_i))}{p(1-p)\pi(\bm{X}_i)(1-\pi(\bm{X}_i))}  \right]$$
where $p$ is the probability of treatment and $\pi(\bm{X}_i)$ is the probability that $S_i=1$ given covariate values $\bm{X}_i$.
\end{proposition}
This can be estimated by the following:
$$\hat{\delta}_{PS} = \frac{1}{N} \sum_{i=1}^N Y_i \frac{(T_i -p)(S_i - \hat{\pi}(\bm{X}_i))}{p(1-p) \hat{\pi}(\bm{X}_i)(1-\hat{\pi}(\bm{X}_i))}$$
where estimation of $\hat{\pi}(\bm{X}_i)$ can be performed using the researcher's preferred propensity score estimation method. Note that this approach is related to the marginal structural model estimation method presented in \cite{vanderweele2009distinction}.

\subsection{Flexible Estimation}

Alternatively, more flexible statistical learning procedures (e.g. random forests) can be employed for estimation of the ATME. Within treatment-level subset $t$, a statistical learning algorithm or procedure can first be employed to learn a flexible function $g_t(S,\bm{X})$, which maps $S$ and $\bm{X}$ to $Y$ and thereby allows for prediction of $Y$ given $S$ and $\bm{X}$. Then for that treatment-level subset, the average causal effect of $S$ on $Y$ can be formulated using partial dependence functions, defined as $g_{St}(S) = E_{\bm{X}} [g_t(S,\bm{X})]$. 

Specifically, the average causal effect of $S$ on $Y$ within treatment-level subset $t$ could be estimated by
$$\bar{g}_{St}(1) - \bar{g}_{St}(0) = \frac{1}{N_t} \sum^{N_t}_{i=1} g_t(1, \bm{x}_{i}) - \frac{1}{N_t} \sum^{N_t}_{i=1} g_t(0, \bm{x}_{i})$$
where $\bm{x}_{i}$ denotes the vector of values in $\bm{X}$ observed in the data for unit $i$. As before, this could be performed separately for each treatment-level subset, $t=0,1$, and the difference between the two would constitute the ATME. 

The precise statistical properties of this estimate of the ATME---consistency, unbiasedness, variance, etc.---would, of course, be subject to the specific properties of the statistical learning procedure(s) being employed.

\clearpage

\section*{Appendix C: Asylum Seeker Application Details}

\setcounter{table}{0}
\setcounter{figure}{0}
\setcounter{subsection}{0}
\renewcommand{\thetable}{C\arabic{table}}%
\renewcommand{\thefigure}{C\arabic{figure}}%

This appendix provides additional details on the application presented in the main text, which is an extension of a previous study \citep{bansaketal2016}. 

\begin{figure}[!hbt]
\caption{Example profile pair}
\subcaption*{\label{fig:example_table}This figure displays an example of a pair of asylum-seeker profiles seen by the survey respondents. This example comes from the English version of the survey administered to respondents in the United Kingdom. Each respondent saw and evaluated five separate pairs. The order of the attributes (rows) was fully randomized between respondents, but for each respondent, the order was kept constant across the five pairs they were shown. The specific attribute levels (values in the cells in the last two columns) were fully randomized between and within respondents. Source: \cite{bansaketal2016}.}
\begin{center}
    \includegraphics[trim = 0 0 0 0, width=1\textwidth]{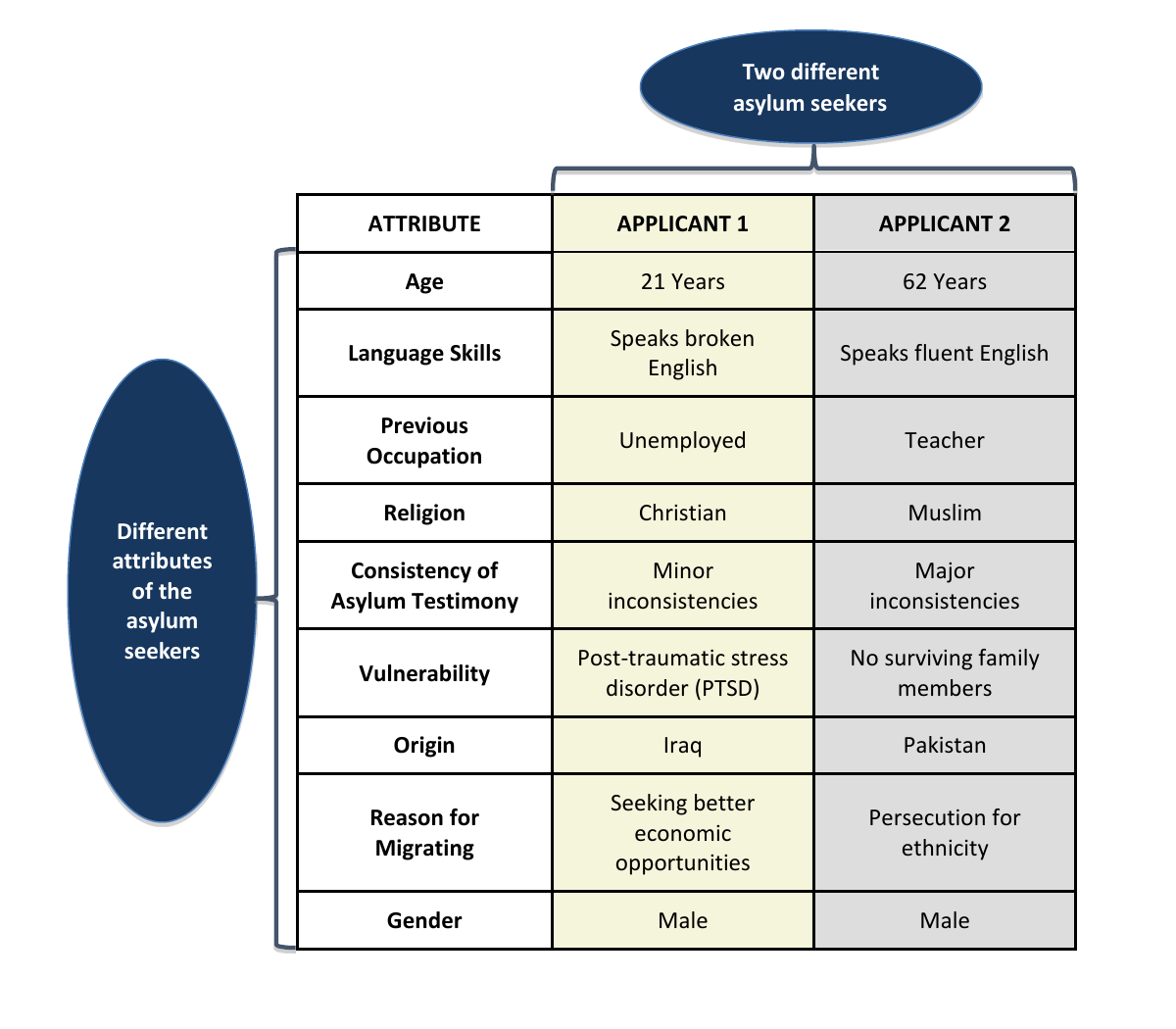}
\end{center}
\end{figure}

Figure \ref{fig:example_table} displays an example of a pair of profiles shown to respondents, and Table \ref{tab:attributes} describes the conjoint attributes. The following is the text used to introduce the respondents to the conjoint section of the survey, as well as the questions used to measure the respondents' evaluations of the asylum-seeker profiles and hence construct the outcome variable. \\

\noindent \textit{Prelude:} \\

\noindent ``Now we would like to show you the profiles of potential applicants for asylum in Europe. You will be shown pairs of asylum seekers, along with several of their attributes. We would like to know your opinion regarding whether you would be in favor of sending each applicant back to their country of origin or allowing them to stay in [Respondent's Country]. \\

\noindent In total, we will show you five comparison pairs. Please take your time when reading the descriptions of each applicant. People have different opinions about this issue, and there are no right or wrong answers." \\

\noindent \textit{Questions:}
\begin{enumerate}
\item \textit{(Rating)} ``On a scale from 1 to 7, where 1 indicates that [Respondent's Country] should absolutely send the applicant back to their country of origin and 7 indicates that [Respondent's Country] should definitely allow the applicant to stay, how would you rate each of the asylum seekers described above?"
\item \textit{(Choice)} ``Now imagine that you had to choose one applicant who would be allowed to stay in [Respondent's Country], and the other applicant would be sent back to their own country of origin. Which of the two applicants would you personally prefer to be allowed to stay in [Respondent's Country]?"
\end{enumerate}

The Choice outcome variable is used for the analysis presented in the main text. In addition to the conjoint component, the survey instrument also contained a number of questions that measured various characteristics and attitudes of the respondents. Tables \ref{tab:variables1}-\ref{tab:sumstat2} describe these additional variables and present summary statistics.

\clearpage

\begin{table}[h!]
\small
\centering
\caption{Conjoint attributes and attribute values/levels}\label{tab:attributes}
\subcaption*{This table displays the attributes and attribute levels used to construct the asylum-seeker profiles.}
\begin{tabular}{ll}
  \toprule
\textbf{Attribute} & \textbf{Attribute Levels} \\
\midrule
\emph{Asylum Testimony} &  No inconsistencies, Minor inconsistencies, Major inconsistencies \\ \\
\emph{Gender} &  Female, Male \\ \\
\emph{Country of Origin} &  Syria, Afghanistan, Kosovo, Eritrea, Pakistan, Ukraine, Iraq \\ \\
\emph{Age} & 21 years, 38 years, 62 years \\ \\
\emph{Previous Occupation} & Unemployed, Cleaner, Farmer, Accountant, Teacher, Doctor \\ \\
\emph{Vulnerability} & None, Post-traumatic stress disorder (PTSD), Victim of torture, \\ 
 & No surviving family members, Physically handicapped \\  \\
\emph{Reason for Migrating} & Persecution for political views, Persecution for religious beliefs, \\ 
 & Persecution for ethnicity, Seeking better economic opportunities \\ \\
\emph{Religion} & Christian, Agnostic, Muslim \\ \\
\emph{Language Skills} & Speaks fluent \emph{[language of respondent's country]}, \\ 
 & Speaks broken \emph{[language of respondent's country]}, \\
 & Speaks no \emph{[language of respondent's country]} \\ \\
\bottomrule
\end{tabular}
\end{table}

\clearpage

\begin{table}[h!]
\small
\centering
\caption{Variables}
\subcaption*{This table describes the outcome variable and other respondent characteristics measured in the survey.} \label{tab:variables1}
\begin{tabular}{ p{2cm} p{14cm}}
  \toprule
\textbf{Variable} & \textbf{Description} \\
\midrule
\emph{Forced Choice} & This is a binary indicator for whether or not a profile was the preferred profile in its respective pair. It is based on respondents' answers to the following question: ``Now imagine that you had to choose one applicant who would be allowed to stay in [Respondent's Country], and the other applicant would be sent back to their own country of origin. Which of the two applicants would you personally prefer to be allowed to stay in [Respondent's Country]?" \\ \\
\emph{Age} & Self-reported age. For the analysis used in the main text, indicators for four age bins were employed: 18-29, 30-44, 45-64, 65+.  \\ \\
\emph{Gender} & Self-reported gender. \\ \\
\emph{Education} & Self-reported highest level of education achieved. The customized country-specific educational level questionnaire options used by the European Social Survey were employed for each country, and the respondents' educational levels were mapped onto the harmonized European version of the International Standard Classification of Education (EISCED) scale. The scale contains seven major categories: less than lower secondary (1), lower secondary (2), lower tier upper secondary (3), upper tier upper secondary (4), advanced vocational/sub-degree (5), lower tertiary education/BA level (6), and higher tertiary education/$\geq$ MA level (7).  \\ \\
\emph{Income} & Self-reported household income. Respondents' household income levels were measured using the customized country-specific household income decile bins constructed in the European Social Survey (the most recent available version of the European Social Survey was used for each country in the sample). For the analysis in the main text, these deciles were coarsened into quintiles. \\ \\
\bottomrule
\end{tabular}
\end{table}

\clearpage

\begin{table}[h!]
\small
\centering
\caption{Variables, continued}
\label{tab:variables2}
\begin{tabular}{ p{2cm} p{14cm}}
  \toprule
\textbf{Variable} & \textbf{Description} \\
\midrule
\emph{Political ideology} & Self-reported placement on the left-right political ideological spectrum. Ideological placement was measured using a standard self-identification scale ranging from 0 on the left to 10 on the right. The question wording used to measure the respondents' ideology was as follows: ``In politics people often talk of `left' and `right'. On this scale from 0 (left) to 10 (right), where would you classify your own political views?" For the analysis in the main text, political ideology was coarsened into five bins: far left (0-2), left (3-4), center (5), right (6-7), and far right (8-10). \\ \\
\emph{Nationalism} &  Multi-construct index of nationalism. To measure nationalism, the survey employed four commonly used prompts---taken from the National Identity module of the International Social Survey Programme \citep{issp2012international} and \cite{mansfield2009support}---tapping into different dimensions of nationalism. For each prompt, respondents specified their level of (dis)agreement: agree strongly (2), agree (1), neither agree nor disagree (0), disagree (-1), disagree strongly (-2). These values were then summed to create an index for nationalism, and the median value was used as a cut point for creating a dichotomized version of the variable (high-versus-low nationalism) for the analysis in the main text. The prompts included the following. (a) ``I would rather be a citizen of [Respondent's Country] than of any other country in the world." (b) ``Generally speaking [Respondent's Country] is a better country than most other countries." (c) ``[Respondent's Country] should follow its own interests, even if this leads to conflicts with other nations." (d) ``[Respondent's Country's] government should just try to take care of the well-being of [Respondent's Country's] citizens and not get involved with other nations."  \\ \\
\bottomrule
\end{tabular}
\end{table}

\clearpage

\begin{table}[h!]
\centering
\caption{Number of respondents per country} \label{tab:respondents}
\begin{tabular}{rr}
  \toprule
 & Sample Size \\
  \midrule
Austria & 1206 \\
  Czech Republic & 1202  \\
  Denmark & 1201  \\
  France & 1203  \\
  Germany & 1200  \\
  Greece & 1200  \\
  Hungary & 1200  \\
  Italy & 1200  \\
  Netherlands & 1200  \\
  Norway & 1202  \\
  Poland & 1201  \\
  Spain & 1203  \\ 
  Sweden & 1203  \\
  Switzerland & 1208  \\ 
  United Kingdom & 1201  \\
   \midrule
  Total & 18030  \\
\end{tabular}
\end{table}

\clearpage

\begin{landscape}

\begin{table}[h!]
\scriptsize
\centering 
\caption{Summary statistics} \label{tab:sumstat1}
\subcaption*{This table displays the distribution of respondents along key demographic and ideological dimensions, by country.}
\begin{tabular}{l|c|ccccc|ccccc|ccc}
\toprule
\multicolumn{1}{l|}{} & \multicolumn{1}{|c|}{Gender} & \multicolumn{5}{|c|}{Age} & \multicolumn{5}{|c|}{Income Quintile} & \multicolumn{3}{|c}{Political Ideology} \\
\midrule
 & Female & 29/under & 30-39 & 40-49 & 50-59 & 60/over & First & Second & Third & Fourth & Fifth & Left & Center & Right \\ 
\midrule
\midrule
Austria & 50\% & 22\% & 19\% & 24\% & 18\% & 17\% & 29\% & 24\% & 22\% & 17\% & 8\% & 34\% & 37\% & 29\% \\ 
  Czech Republic & 51\% & 42\% & 28\% & 16\% & 9\% & 5\% & 15\% & 19\% & 23\% & 28\% & 15\% & 22\% & 43\% & 35\% \\ 
  Denmark & 50\% & 20\% & 21\% & 22\% & 20\% & 17\% & 26\% & 20\% & 25\% & 16\% & 13\% & 37\% & 25\% & 38\% \\ 
  France & 51\% & 23\% & 20\% & 21\% & 20\% & 15\% & 29\% & 23\% & 17\% & 23\% & 9\% & 30\% & 35\% & 36\% \\ 
  Germany & 50\% & 20\% & 18\% & 25\% & 21\% & 16\% & 25\% & 23\% & 21\% & 20\% & 11\% & 41\% & 36\% & 24\% \\ 
  Greece & 47\% & 33\% & 29\% & 25\% & 10\% & 3\% & 37\% & 28\% & 19\% & 11\% & 5\% & 36\% & 36\% & 28\% \\ 
  Hungary & 51\% & 43\% & 28\% & 14\% & 8\% & 6\% & 18\% & 20\% & 15\% & 21\% & 26\% & 19\% & 48\% & 33\% \\ 
  Italy & 50\% & 19\% & 22\% & 23\% & 19\% & 17\% & 27\% & 26\% & 22\% & 18\% & 7\% & 35\% & 26\% & 39\% \\ 
  Netherlands & 50\% & 22\% & 17\% & 23\% & 21\% & 17\% & 30\% & 22\% & 19\% & 19\% & 9\% & 28\% & 33\% & 39\% \\ 
  Norway & 49\% & 22\% & 22\% & 21\% & 19\% & 16\% & 28\% & 26\% & 19\% & 13\% & 13\% & 34\% & 24\% & 42\% \\ 
  Poland & 49\% & 19\% & 30\% & 25\% & 15\% & 11\% & 15\% & 22\% & 20\% & 24\% & 20\% & 23\% & 39\% & 38\% \\ 
  Spain & 50\% & 22\% & 25\% & 22\% & 17\% & 14\% & 23\% & 25\% & 18\% & 16\% & 17\% & 51\% & 23\% & 27\% \\ 
  Sweden & 49\% & 22\% & 20\% & 21\% & 19\% & 18\% & 22\% & 15\% & 20\% & 18\% & 25\% & 37\% & 26\% & 37\% \\ 
  Switzerland & 50\% & 22\% & 20\% & 23\% & 19\% & 16\% & 33\% & 29\% & 20\% & 14\% & 4\% & 29\% & 30\% & 41\% \\ 
  United Kingdom & 52\% & 26\% & 19\% & 21\% & 18\% & 16\% & 29\% & 21\% & 19\% & 19\% & 12\% & 27\% & 43\% & 30\% \\ 
\bottomrule
\end{tabular}
\end{table}

\clearpage

\begin{table}[h!]
\scriptsize
\centering
\caption{Summary statistics, continued} \label{tab:sumstat2}
\subcaption*{``Lower Secondary and Below" corresponds to EISCED 1-2, ``Upper Secondary" EISCED 3-4, ``Advanced Vocational" EISCED 5, ``BA Level" EISCED 6, and ``MA Level and Above" EISCED 7.} 
\begin{tabular}{l|ccccc}
  \toprule
 \multicolumn{1}{l|}{} & \multicolumn{5}{|c}{Education} \\
 \midrule
 & Lower & Upper & Advanced & BA Level & MA Level \\ 
  & Secondary &  Secondary &  Vocational & &  and Above \\ 
   & and Below &   &   & &   \\ 
 \midrule
\midrule
Austria & 8\% & 53\% & 20\% & 5\% & 13\% \\ 
  Czech Republic & 5\% & 56\% & 15\% & 10\% & 14\% \\ 
  Denmark & 22\% & 37\% & 10\% & 22\% & 8\% \\ 
  France & 10\% & 45\% & 21\% & 9\% & 16\% \\ 
  Germany & 5\% & 38\% & 29\% & 15\% & 13\% \\ 
  Greece & 4\% & 32\% & 19\% & 29\% & 16\% \\ 
  Hungary & 4\% & 49\% & 18\% & 19\% & 9\% \\ 
  Italy & 13\% & 48\% & 4\% & 11\% & 23\% \\ 
  Netherlands & 32\% & 34\% & 7\% & 12\% & 15\% \\ 
  Norway & 13\% & 35\% & 17\% & 23\% & 12\% \\ 
  Poland & 9\% & 35\% & 13\% & 14\% & 29\% \\ 
  Spain & 23\% & 19\% & 16\% & 17\% & 25\% \\ 
  Sweden & 12\% & 38\% & 28\% & 11\% & 11\% \\ 
  Switzerland & 10\% & 50\% & 24\% & 7\% & 9\% \\ 
  United Kingdom & 17\% & 37\% & 14\% & 21\% & 11\% \\ 
 \bottomrule
\end{tabular}
\end{table}

\end{landscape}

\clearpage

\section*{Appendix D: Sensitivity Analysis}

\setcounter{table}{0}
\setcounter{figure}{0}
\setcounter{subsection}{0}
\renewcommand{\thetable}{D\arabic{table}}%
\renewcommand{\thefigure}{D\arabic{figure}}%

This appendix shows how sensitivity analysis can be applied to estimation of the ATME using the parallel estimation framework. More specifically, it 
describes an adaptation of a sensitivity analysis developed by \citet{imbens2003}. For illustrative purposes, it applies the adapted sensitivity analysis to an auditing and corruption field experiment \citep{olken2007}.

\subsection{Background on \cite{olken2007}}

Recent research in economics and political science has increasingly used experimental designs to study the effectiveness of various policies to curb undesirable behavior by elites and politicians, such as engagement in corruption and clientelism. One well-known example is a study by \citet{olken2007} on the effects of auditing on corruption in construction projects in Indonesia. In one of his analyses, Olken investigates how having one's village construction project be randomly assigned to an auditing treatment affected a household's probability of having a member hired to work on the project. While Olken shows (in a separate analysis) that the audit treatment decreases the misappropriation of project funds, he theorizes that this decrease in one type of corruption may be accompanied by the increase in a different type, namely nepotism. To investigate this, he tests whether having a family member as the construction project head moderates (namely, increases) the audit treatment effect on the probability that someone in the household was hired to work on the project.

As is standard in the literature, Olken employs a controlled interaction regression to estimate this moderation effect, regressing the outcome on the treatment, moderator, and treatment-moderator interaction, as well as including a wide range of control variables but not interacting them with the treatment. The estimated interaction in this specification between the audit treatment dummy and a dummy indicator for having a family member as the construction project head is $0.138$, suggesting that having a family member as the construction project head increases the effect of the audit treatment on the outcome by approximately 14 percentage points.\footnote{Olken also includes a specification that includes interactions between the treatment and two other possible moderator variables in addition to the interaction discussed here, but it does not include interactions between the treatment and the full range of potential confounding variables.} If the parallel regression approach presented in this study (or equivalently, a single regression including interactions between the treatment and all possible confounding variables) is used, the moderation effect is attenuated to $0.120$, but remains statistically significant.

\subsection{Background on \cite{imbens2003}}

The sensitivity analysis developed by \cite{imbens2003} in its original form involves a parametric model in which several simplifying modeling assumptions are made. First, the outcome variable $Y$ is assumed to follow a normal distribution with a mean that is a function of a non-exogenous binary treatment $T$, observed confounders $\bm{X}$, and an unobserved confounder $U$. Second, given $U$ and $\bm{X}$, the probability of the treatment $T$ is modeled as following a logistic distribution. Third, the unobserved confounder $U$ is modeled as a Bernoulli random variable that is, without loss of generality, independent of $\bm{X}$. By making these modeling assumptions, Imbens specifies a likelihood function whereby maximum likelihood estimation (MLE) can be used to obtain the hypothetical estimates of the treatment effect had the hypothetical confounder $U$ been observed. These estimates depend upon specified relationships, in the form of parameter values fixed by the analyst, between $U$ and $Y$ (effect of $U$ on $Y$) and between $U$ and $T$ (selection effect into $T$ given $U$). Using subject matter expertise combined with parameter estimates involving observed confounders, the analyst can choose plausible ranges of parameters to model the confounding by $U$ and, under those hypothetical conditions, estimate the extent to which the estimated treatment effect would change---i.e. its sensitivity to unobserved confounding.

\subsection{Adapting the Sensitivity Analysis}

The version of this sensitivity analysis extended from the ATE to ATME context, developed and presented here, proceeds in the same manner with two major differences. First, instead of including the treatment in the parametric model described above, the moderator is included in its place. This is because the scenario to which the parallel estimation framework pertains is one in which the treatment is randomized/exogenous but the moderator is not. Thus, it is potential unobserved confounding of the moderator, not the treatment, that must be modeled. Second, in accordance with the parallel estimation process, the full parametric model and subsequent estimation via MLE is performed separately for the treatment and control subsets.

Formally, for treatment $T$, moderator $S$, potential outcome $Y(s)$, control variable vector $\bm{X}$, and hypothetical unobserved confounder $U$, the following model is applied separately for subsets $T=0,1$:
\begin{eqnarray} \nonumber
U & \sim & B(1,1/2) \\ \nonumber
Pr(S = 1 | \bm{X}, U) & = & \frac{exp(\zeta + \bm{X}' \bm{\eta} + \alpha U)}{1+exp(\zeta + \bm{X}' \bm{\eta} + \alpha U)} \\ \nonumber
Y(s) | \bm{X},U & \sim & N(\xi + \gamma s + \bm{X}' \bm{\beta} + \kappa U, \sigma^2)
\end{eqnarray}
As in the standard version of the Imbens sensitivity analysis, the parameters relating the hypothetical unobserved confounder $U$ with the moderator and outcome ($\alpha$ and $\kappa$ respectively) are fixed to certain plausible values specified by the analyst in accordance with their (possibly worst-case) expectations. In the standard version of the Imbens sensitivity analysis, because the model is applied to only one sample of data, only two parameters must be specified. In contrast, in the version presented here, there are four parameters to be fixed since there is an $\alpha$ and $\kappa$ for each treatment subset.

The enlargement of the fixed parameter set may, at first consideration, appear to overcomplicate this version of the sensitivity analysis. However, this turns out not to be the case. First, given Assumptions \ref{assump:exog} and \ref{assump:pretre} that the treatment is randomized and that the moderator and confounding variables (including $U$) are not affected by the treatment, the selection effect into $S$ given $U$ (i.e. $\alpha$) is equal for both the treatment and control subsets. Hence, when fixing the $\alpha \equiv \tilde{\alpha}$ values in the sensitivity analysis, $\tilde{\alpha}$ should be set to the same value for both subsets $T = 0,1$, reducing as before to a single fixed parameter. In contrast, the same cannot be done for $\kappa$, the effect of $U$ on $Y$. For there to be moderation of the treatment effect by $S$, then it is also the case that $S$ affects $Y$ differently depending upon whether $T=0$ or $T=1$. Similarly, the confounding of the estimate of the ATME by an unobserved variable is due precisely to the possibility that the unobserved confounding variable also affects $Y$ differently depending upon whether $T=0$ or $T=1$. In other words, we are concerned that the ATME estimate is confounded because an unobserved variable also moderates the treatment effect. Further, just as $\gamma_{T=1} - \gamma_{T=0}$ represents the ATME of interest, $\kappa_{T=1} - \kappa_{T=0}$ represents the analogous moderation of the treatment effect by $U$. As a result, while $\kappa_{T=1}$ and $\kappa_{T=0}$ must be fixed separately, the sensitivity analysis should focus on their difference, reducing the task of interpretation, if not the underlying parameters, to a single value of interest.

In sum, this sensitivity analysis allows for the investigation of how sensitive the ATME estimate is (e.g. the extent to which it would be attenuated toward zero) given two primary considerations: a fixed hypothetical selection effect into the moderator given $U$ and a fixed hypothetical moderation effect on the treatment caused by $U$. Formally, for fixed $\tilde{\alpha}$ and $\tilde{\kappa}_{T=1} - \tilde{\kappa}_{T=0}$, the implied estimate of the ATME, $\widetilde{\widehat{\delta}} = \tilde{\hat{\gamma}}_{T=1} - \tilde{\hat{\gamma}}_{T=0}$, can be computed via MLE.

\begin{figure}[ht!]
\begin{center}
\caption{Level Curve Depicting the Strength of the Unobserved Confounder's Relationships with Treatment and Moderator Required for ATME Estimate to Halve in Size}\label{fig:sens}
\includegraphics[scale=0.7]{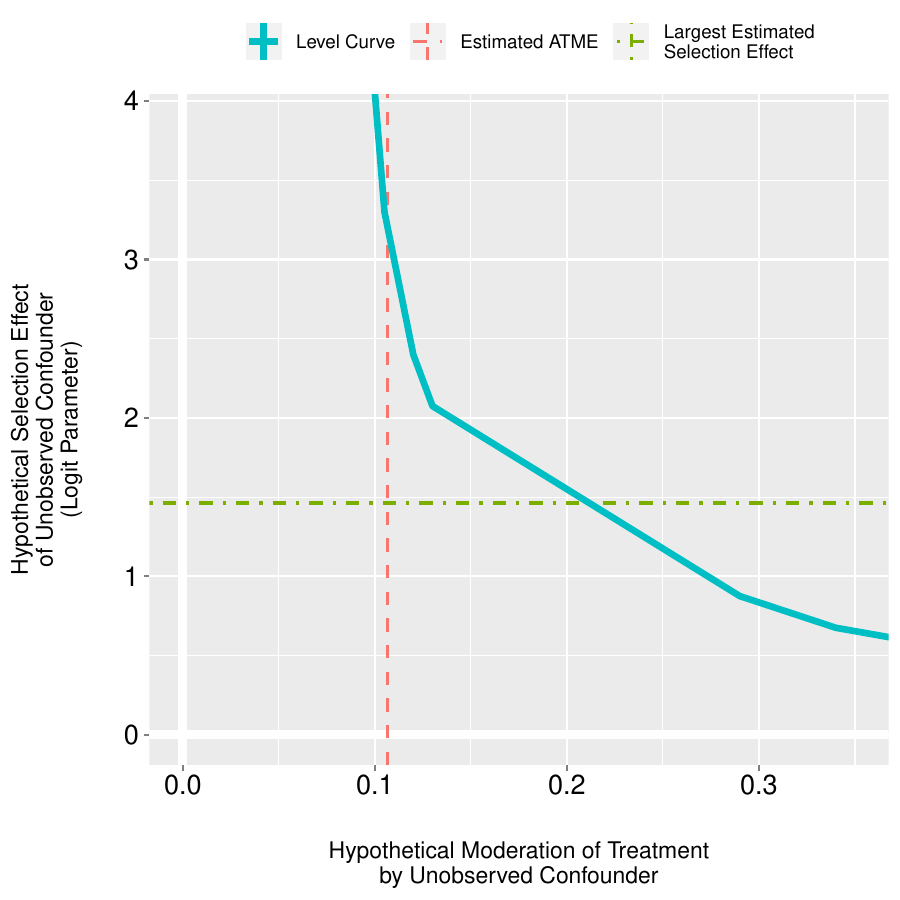}
\end{center}
\end{figure}

To illustrate, the adapted sensitivity analysis is applied to the \citet{olken2007} experimental data on the effects of auditing on corruption in construction projects in Indonesia. As described earlier, Olken tests whether having a family member as the head of the construction project moderates (namely, increases) the audit treatment effect on the probability of having a household member hired to work on the project. While this turns out to be a case in which specifying the proper model for estimating causal moderation does not dramatically change the resulting estimate relative to conventional approaches, there is still the possibility of the ATME estimate being confounded by unobserved variables. Thus, the sensitivity analysis as described above is applied to the Olken data and model.\footnote{For computational efficiency purposes, the stratum fixed effects included in Olken's model are dropped. These stratum fixed effects have little effect on the estimates in the original models, thus suggesting that dropping them is unlikely to seriously affect the inferences drawn from this analysis.} To enable easier inspection and visualization of the sensitivity of the causal moderation effect, the implied $\widetilde{\widehat{\delta}}$ was estimated over a grid of plausible $\tilde{\alpha}$ and $\tilde{\kappa}$ values. Figure \ref{fig:sens} displays a set of results from the sensitivity analysis. 

Under consideration in the figure is what strength of unobserved confounding would be required to cut the estimated ATME in half (though any value of interest could be specified for this). The $y$-axis measures the hypothetical selection effect into the moderator attributable to the unobserved confounder ($\tilde{\alpha}$), while the $x$-axis measures the hypothetical causal moderation of the treatment by the unobserved confounder ($\tilde{\kappa}_{T=1} - \tilde{\kappa}_{T=0}$). The blue curve is a level curve depicting the combination of values for these two effects required to halve the estimated ATME. The vertical red and horizontal green dashed lines provide reference points for interpreting the extent of confounding implied by the level curves. The horizontal green dashed line marks the largest (in absolute value) estimated selection effect by a variable that is observed in the data. This variable is closely related to the moderator (whereas the moderator is having a family member as project head, this other variable is an indicator for having a family member in the village government). Further, it is also a binary variable, like the unobserved confounder, thus providing a clear theoretical reference point for how strong of a relationship is implied by $\tilde{\alpha}$ values above the green line. The vertical red dashed line marks the parallel regression estimate of the ATME, thus providing a clear reference point for the hypothetical causal moderation effect of the unobserved confounder.

In sum, if the true ATME is only half the size of the estimated ATME due to estimation bias caused by an unobserved confounder, the blue level curve depicts how significant that confounder would need to be in terms of its relationship with the moderator of interest and its own impact on the treatment effect. Because the blue level curve does not pass through the lower-left box formed by (below and to the left of) the green and red lines---which may be considered a sensitivity danger zone---the results of this sensitivity analysis can be interpreted as evidence that it is unlikely for there to exist an unobserved confounder significant enough to imply an ATME that is only half the size of the estimated ATME. Thus, while the existence of unobserved confounders is virtually certain, we can be relatively confident in the existence of a causal moderation effect of the sort posited by Olken.

\end{document}